# Quadrupolar Excitons and Hybridized Interlayer Mott Insulator in a Trilayer Moiré Superlattice

Zhen Lian[1#], Dongxue Chen[1#], Lei Ma[1#], Yuze Meng[1#], Ying Su[2], Li Yan[1], Xiong Huang[3,4], Qiran Wu[3], Xinyue Chen[1], Mark Blei[5], Takashi Taniguchi[6], Kenji Watanabe[7], Sefaattin Tongay[5], Chuanwei Zhang[2], Yong-Tao Cui[3*], Su-Fei Shi[1,8*]

1. Department of Chemical and Biological Engineering, Rensselaer Polytechnic Institute, Troy, NY 12180, USA
2. Department of Physics, University of Texas at Dallas, Dallas, Texas, 75083, USA
3. Department of Physics and Astronomy, University of California, Riverside, California, 92521, USA
4. Department of Materials Science and Engineering, University of California, Riverside, California, 92521, USA
5. School for Engineering of Matter, Transport and Energy, Arizona State University, Tempe, AZ 85287, USA
6. International Center for Materials Nanoarchitectonics, National Institute for Materials Science, 1-1 Namiki, Tsukuba 305-0044, Japan
7. Research Center for Functional Materials, National Institute for Materials Science, 1-1 Namiki, Tsukuba 305-0044, Japan
8. Department of Electrical, Computer & Systems Engineering, Rensselaer Polytechnic Institute, Troy, NY 12180, USA

[#] These authors contributed equally to this work
[*] Corresponding authors: yongtao.cui@ucr.edu, shis2@rpi.edu

## Abstract

**Transition metal dichalcogenide (TMDC) moiré superlattices, owing to the moiré flatbands and strong correlation, can host periodic electron crystals and fascinating correlated physics. The TMDC heterojunctions in the type-II alignment also enable long-lived interlayer excitons that are promising for correlated bosonic states, while the interaction is dictated by the asymmetry of the heterojunction. Here we demonstrate a new excitonic state, quadrupolar exciton, in a symmetric $WSe_2$-$WS_2$-$WSe_2$ trilayer moiré superlattice. The quadrupolar excitons exhibit a quadratic dependence on the electric field, distinctively different from the linear Stark shift of the dipolar excitons in heterobilayers. This quadrupolar exciton stems from the hybridization of $WSe_2$ valence moiré flatbands. The same mechanism also gives rise to an interlayer Mott insulator state, in which the two $WSe_2$ layers share one hole laterally confined in one moiré unit cell. In contrast, the hole occupation probability in each layer can be continuously tuned via an out-of-plane electric field, reaching 100% in the top or bottom $WSe_2$ under a large electric field, accompanying the transition from quadrupolar excitons to dipolar excitons. Our work demonstrates a trilayer moiré system as a new exciting playground for realizing novel correlated states and engineering quantum phase transitions.**

## Introduction

Monolayer TMDCs, as atomically thin direct bandgap semiconductors, offer a unique playground to explore novel optoelectronic phenomena[1,2], especially with the ability to form heterostructures that enable a new range of control knobs. For example, TMDC heterojunctions in a type-II alignment host long-lived interlayer excitons[3–6], with electrons and holes residing in different layers[3,4]. These interlayer excitons possess the valley degree of freedom, as well as a large Stark shift due to the permanent dipole moment, rendering them promising candidates as tunable quantum emitters[6]. Recently, angle-aligned TMDC moiré superlattices exhibit strong Coulomb interactions in the electronic flatbands, leading to correlated states[7–19] such as Mott insulator and generalized Wigner crystal[7,8,12,17]. The moiré coupling also gives rise to flat excitonic bands[20–23] that could potentially be utilized to realize correlated bosonic states[24], such as Bose-Einstein condensation (BEC) and superfluidity[25–27]. The interaction between interlayer excitons is dominated by the repulsive force between their permanent dipoles, whose alignment is dictated by the asymmetry of the heterostructure, with electrons and holes separated in two different layers.

In this work, we report a new interlayer quadrupolar exciton in a symmetric TMDC heterostructure: angle-aligned $WSe_2/WS_2/WSe_2$ trilayer. The interlayer excitons in the top and bottom bilayers have opposite polarities, which restores the symmetry. Their hybridization then forms a superposition state of interlayer excitons, canceling the dipolar moments and giving rise to a quadrupolar exciton, which has been predicted to enable intriguing quantum phase transition[26,28–30]. In the presence of moiré coupling, this hybridization further gives rise to a new type of correlated electronic state, hybridized interlayer Mott insulator, in which the correlated holes are shared between the two $WSe_2$ layers, and the layer population can be continuously tuned by an electric field.

## Results and Discussion

### Quadrupolar Exciton in Angle-Aligned $WSe_2/WS_2/b$-$WSe_2$ Trilayer

The typical device structure is schematically shown in Fig. 1a, which contains three regions of different stackings among the three monolayers: (I) $WS_2$ over the bottom $WSe_2$, which we denote as $WS_2/b$-$WSe_2$; (II) top $WSe_2$ over $WS_2$ (t-$WSe_2/WS_2$); and (III) the t-$WSe_2/WS_2/b$-$WSe_2$ trilayer. The whole device is gated by the top and bottom gate electrodes made of few-layer graphene (FLG), which provide independent control of the electric field and doping.

In the bilayer regions I and II, the $WSe_2/WS_2$ moiré superlattices host both correlated electrons and interlayer excitons due to the type-II band alignment (Fig. 1c). The interlayer excitons, with holes residing in the $WSe_2$ layer and electrons in the $WS_2$ layer (Fig. 1c), interact with the correlated electrons and can be used to read out the transitions at the correlated insulating states[17,31–34]. The doping-dependent photoluminescence (PL) spectra in these regions (Figs. 1e and 1f) clearly reveal these features: the interlayer exciton PL peak has a strong intensity at the charge neutrality point (CNP), which decreases quickly upon doping; the PL energy and intensity are also modulated by correlated insulator states such as the Mott insulator states at both n=1 and -1, consistent with the previous studies[12,31].

In the trilayer region III, we expect quadrupolar excitons as schematically plotted in Fig. 1b. The quadrupolar exciton is the superposition of the two dipolar excitons of opposite polarities through the hybridization of the valence bands in the top and bottom $WSe_2$ layers, which leads to the splitting of valence bands, $\Delta^{\pm}$, as shown in Fig. 1d, similar to the formation of bonding and antibonding states in a double-well system[29]. As a result, the quadrupolar excitons will have two branches: one at lower energy than the dipolar exciton and the other at higher energy, assuming that all have similar binding energies[29]. Fig. 1g plots the PL in this region, which indeed exhibits a major PL resonance at energies below the dipolar excitons in Figs. 1e and 1f. We have not observed any PL resonance corresponding to the higher energy quadrupolar exciton yet, while some devices show high energy exciton PL with different nature that we are going to explore in the future (details in Supplementary Information Section 18). The doping dependence is also drastically different: the intensity of the lower energy PL peak retains upon hole doping and only starts to decay at n=-1 (we will discuss this in more detail later).

The quadrupole nature of the excitons in the trilayer region is confirmed by the electric field dependent PL spectra. In regions I (Fig. 2a) and II (Fig. 2b), the interlayer exciton PL peaks both shift linearly as a function of the out-of-plane electric field but with opposite signs of the slope. The slope is -0.72 $e \cdot nm$ for $WS_2$/b-$WSe_2$ (region I) and 0.66 $e \cdot nm$ for t-$WSe_2$/$WS_2$ (region II), consistent with the previous results[21,35–39]. In contrast, the PL from the trilayer region III is symmetric about the electric field, and the resonance energy exhibits a quadratic dependence on the electric field, as shown in Figs. 2c and 2d, clearly demonstrating that the trilayer PL is from quadrupolar excitons. The PL resonance energy can be well fitted by a quadrupolar exciton model (orange curves in Fig. 2d, details in Supplementary Information Section 9). It is worth noting that at large electric fields, the quadrupolar exciton approaches the linear Stark shift of dipolar excitons with a slope around 0.7 $e \cdot nm$ (dashed lines), matching what we extracted from the data in the bilayer regions I and II. We further extract the $\Delta^{DQ}$, the energy difference between dipolar excitons and quadrupolar excitons under net zero electric field, to be about 12 meV from the fitting in Fig. 2d (Supplementary Information Section 9), consistent with the theoretical calculation for a similar trilayer structure (10-30 meV in $WSe_2$/$MoSe_2$/$WSe_2$)[29]. We have also reproduced similar quadrupolar exciton behaviors in other angle-aligned $WSe_2$/$WS_2$/$WSe_2$ devices, which show a $\Delta^{DQ}$ about 30 meV (device D2, Supplementary Information Section 10) and 9 meV (device D1 and D3, Supplementary Information Section 14). We note that the dipolar exciton resonance energies in regions I and II only serve as a guide of the two dipolar excitons involved in forming the quadrupolar excitons due to dielectric environment difference and possible spatial inhomogeneity. The energies of the two dipolar excitons that form quadrupolar excitons in region III can be extracted from the fitting and are similar in values, typically less than 7 meV (detailed discussion in Supplementary Information Section 14). In fact, the electric field dependence of the quadrupolar exciton can be used to extract the energy difference between the two dipolar excitons involved in the hybridization, which dictates the hybridization to occur at a finite electric field that tunes the two dipolar exciton energies into resonance (details in Supplementary Information Section 14). We also want to mention that the higher energy mode of the predicted quadrupolar excitons (asymmetric quadrupolar exciton mode[29]) is missing in Fig. 2, likely due to the excited state or even dark state nature[40] of the quadrupolar exciton, which leads to the absence of PL.

The quadrupolar excitons show distinctively different power dependence compared with that from dipolar excitons, as shown in Fig. S5. The integrated PL intensity of quadrupolar excitons exhibits more nonlinear dependence than dipolar excitons, likely due to their larger size. In addition, the PL peak blueshifts as a function of the excitation power (Figs. S5b, e) or exciton density (Fig. S6) is smaller for quadrupolar exciton compared with that of dipolar excitons, consistent with our expectation of reduced exciton-exciton repulsion for quadrupolar excitons. The estimation of the exciton density can be found in Supplementary Information Section 16.

## Evidence of An Interlayer Mott Insulator

Next, we study the interaction between the quadrupolar exciton and the correlated electrons in the moiré flatlands. We first revisit the doping dependence of the quadrupolar exciton at zero electric field. Here, the filling factor denotes the number of holes per moiré unit cell ("-" sign for holes), the same as those in the moiré bilayer regions I and II. However, since the trilayer consists of two moiré superlattices, both of which can be filled with carriers, we define their individual filling factors as $n_t$ and $n_b$, respectively, and the total filling factor $n=n_t+n_b$. We focus on the low energy mode of the trilayer quadrupolar exciton and observe two main features in its PL spectra, at $n=-1$ and $n=-2$, respectively. At $n=-1$, the PL peak energy exhibits a kink (Fig. 3b), and the PL intensity drops sharply upon further hole doping (Fig. 3c). At $n=-2$, the PL energy exhibits a blueshift. These features correspond to the emergence of insulating states, similar to the previous studies[17,31–34].

The behaviors at these two fillings evolve systematically as a function of the external electric field. Since the device structure is symmetric, the observed PL behaviors are also symmetric with respect to the electric field direction. Fig. 3d-3g plot examples of PL spectra at selected negative electric fields (direction definition in Fig. 1a), while detailed data at both electric field directions are available in Supplementary Information Section 8. Note that the labeled values of external electric fields are calculated based on voltages applied on the top and bottom gates (see Methods). The effective electric fields between the top and bottom $WSe_2$ layers will be different due to carrier populations and layer chemical potentials (Supplementary Information Section 11). As an electric field is applied, the PL spectra of the low energy quadrupolar mode remain largely unchanged concerning the two main features described above in the low field regime (Figs. 3d and 3e). However, it changes drastically at high electric fields (Figs. 3f and 3g): the PL intensity drops quickly when doped away from CNP, and the PL energy exhibits a blueshift at $n=-1$ instead of

n=-2. In fact, the PL spectra at high electric fields resemble that of dipolar excitons in a moiré bilayer (Figs. 1e and 1f, as well as our previous study[26]). Therefore, the observed change in the PL spectra signals the transition from a quadrupolar exciton to a dipolar exciton. Similar results were reproduced in another device with the same structure (device D3), as shown in Supplementary Information Section 12.

With the understanding of the quadrupolar to dipolar exciton transition, we now discuss the nature of the n=-1 and -2 states and their evolution under electric fields. Figs. 4a and 4c plot the PL intensity and peak energy as a function of both doping (filling factor) and external electric field, respectively. At n=-1, the PL energy and intensity both change abruptly above a certain threshold external electric field $E_{c,-1}$ (about 44 mV/nm), while at n=-2, the PL blueshift disappears when the external electric field exceeds $E_{c,-2}$ (about 32mV/nm). For the n=-1 state, in the absence of an external electric field, each hole is hybridized between the top and bottom WSe$_2$ layers with equal probability, i.e., the hole wavefunction is a superposition of the top and bottom WSe$_2$ valence moiré bands. Laterally it is confined such that there is one hole in the two overlapping moiré unit cells combined (Fig. 4d). This state is a new type of correlated state in the trilayer moiré superlattice, an interlayer Mott insulator. The hole is allowed to tunnel between the top and bottom WSe$_2$ layers in the overlapping moiré cells, but tunneling to neighboring moiré cells is prohibited by the strong Coulomb repulsion. As the electric field increases, for example, in the positive direction defined in Fig. 1a, the probability of holes in the bottom WSe$_2$ layer will increase. Above the threshold electric field $E_{c,-1}$, the hole will be 100% in the bottom WSe$_2$ layer (n$_b$=-1), leaving the top WSe$_2$ layer empty (n$_t$=0). This state now becomes a Mott insulator in the WS$_2$/b-WSe$_2$ interface only, similar to that in a WS$_2$/WSe$_2$ moiré bilayer. The system should remain insulating, as seen in the behavior of the PL peak energy in Fig. 4c. This transition is the result of the competition between the interlayer and intralayer hopping processes, which we characterize as energy $t'$ and $t$, respectively. The interlayer (intralayer) hopping favors carriers populating both (individual) WSe$_2$ layers. Based on the threshold electric field, we estimate the overall potential difference between the two WSe$_2$ layers is about 0 meV at the transition, which suggests that $t'$ is about the same as $t$ (See Supplementary Information Section 11: case 2 for a detailed discussion). We note that it is critical to have similar twist angles to observe the reported hybridized Mott insulator state here. The small difference in the twist angles of the reported device might lead to a moiré superlattice of a much larger period, which is not likely to affect our experimental observation due to the corresponding low density of carriers for the half-filling.

At n=-2, the transition is different. Initially, at zero field, there is one hole per moiré unit cell in each of the two WSe$_2$ layers, forming two separate Mott insulator states at both t-WSe$_2$/WS$_2$ and WS$_2$/b-WSe$_2$ interfaces (Fig. 4e). Application of an electric field will create an energy shift between the two Mott insulator Hubbard bands. However, since both upper Hubbard bands (UHB) are fully occupied by holes, tunneling of holes between the two layers is forbidden, and this carrier configuration (n$_t$=n$_b$=-1) will remain stable until the UHB of the top WSe$_2$ layer starts to overlap with the lower Hubbard band (LHB) of the bottom $WSe_2$ layer (Fig. 4g), and holes from this top layer UHB will start to move to the LHB in the bottom $WSe_2$ layer, resulting in partially filled bands in both layers such that the system will no longer be insulating (see the n=-2 evolution in Fig. 4c). The energy difference between the two $WSe_2$ layer at the transition should be equal to the difference between the onsite Coulomb repulsion, U, and $t' - t$. This potential difference is estimated to be ~20 meV from the threshold field. As $t' - t$ is about 0, this suggests a value of about 20 meV for U, consistent with the previous studies[12,41]. We note that the threshold electric field at n=-2 has a large uncertainty due to the weak PL signals, and the resulting estimation of U is a lower bound.

Finally, the temperature-dependent PL spectra (Fig. S7) show that the interlayer Mott insulator transition temperature is about 80 K, consistent with our expectation based on previous studies on Mott insulator state in $WS_2$/$WSe_2$ moiré systems[41,42]. The quadrupolar excitons, however, are still obvious at 100 K.

We note that we have also observed quadrupolar excitons and correlated states in $WS_2$/$WSe_2$/$WS_2$ trilayer moiré devices in which the conduction bands in the two $WS_2$ layers are hybridized (Supplementary Information Section 19). We choose to focus on the $WSe_2$/$WS_2$/$WSe_2$ trilayer system in this work as the hybridization and interlayer Mott insulator only involve one valence band in each $WSe_2$ monolayer instead of two conduction bands in each $WS_2$ monolayer, which simplifies the system.

In summary, our study demonstrates a unique trilayer moiré system that hosts both quadrupolar excitons and correlated states at n=-1 (interlayer Mott insulator) and n=-2 (Mott insulator). In particular, the quadrupolar excitons and interlayer Mott insulator both originate from the valence band hybridization and interact with each other. Here, the flat valence band hybridization, combined with the large spin-orbit coupling, is promising for generating nontrivial topological states and engineering quantum states such as quantum anomalous Hall[43]. The quadrupolar excitons in this unique trilayer moiré system are not only promising for realizing the quantum phase transition of bosonic quasiparticles but also strongly interact with correlated electrons, setting up an exciting platform for engineering new correlated physics of fermions, bosons, and a mixture of both[44]. We also envision that further development in aligning the moiré trilayer to allow different stacking of moiré sites (high symmetry points[45]) such as AAA, ABA, or ABC will usher in unprecedented opportunities in electronic and excitonic band engineering.

Note: During the submission of this work, we became aware of other works on quadrupolar excitons (Ref.[46], Ref.[47], and reference 29 in Ref.[47] ).

## Method

### Sample fabrication

We used the same dry pick-up method as reported in our earlier work[32] to fabricate TMDC heterostructures. The gold electrodes are pre-patterned on the Si/SiO$_2$ substrate. The monolayer TMDC flakes, BN flakes, and few-layer graphene (FLG) flakes are exfoliated on silicon chips with 285 nm thermal oxide. It is worth noting that typical large TMDC flakes with one dimension exceeding 50 μm were chosen for the device structure shown in Fig. 1. The polycarbonate (PC)/ polydimethylsiloxane (PDMS) stamp was used to pick up TMDC monolayer and other flakes sequentially. The top WSe$_2$ and bottom WSe$_2$ are aligned with a 0-twist angle (R-stacked configuration). This is achieved either through angle-aligned layer stacking and checking the second harmonic generation (SHG) afterward or using the same WSe$_2$ flake and splitting it into two pieces via the tear and twist technique. The alignment of each layer is achieved under a home-built microscope transfer stage with the rotation controlled with an accuracy of 0.02 degrees. The PC is then removed in the chloroform/isopropanol sequence and dried with nitrogen gas. The final constructed devices were annealed in a vacuum (<10$^{-6}$ torr) at 250 °C for 8 hours.

### Optical characterizations.

During the optical measurements, the sample was kept in a cryogen-free optical cryostat (Montana Instruments). A home-built confocal imaging system was used to focus the laser onto the sample (with a beam spot diameter ∼ 2 μm) and collect the optical signal into a spectrometer (Princeton Instruments). During the measurements, the samples were kept in a vacuum and cooled down to 6-10 K. The PL measurements in Figs. 1 and 2 are performed with 50 μW 633 nm CW excitation. All other PL measurements were performed with 633 nm CW excitation with a power of 200 μW unless specified. The reflectance contrast measurements were performed with a super-continuum laser (YSL Photonics). The polarized SHG measurements were performed with a pulsed laser excitation centered at 900 nm (Ti: Sapphire; Coherent Chameleon) with a repetition rate of 80 MHz and a power of 80 mW. The angle between the laser polarization and the crystal axes of the sample was fixed. The SHG signal was analyzed using a half-waveplate and a polarizer. Additional PL measurements were performed with a 730 nm CW diode laser (Supplementary Information Section 15), which showed similar results as the main text.

### Calculation of electric field

The external electric field is defined as $\frac{1}{2}(\frac{V_{TG}}{d_1} - \frac{V_{BG}}{d_2})$, where $V_{TG}$ (($V_{BG}$) is the top (back) gate voltage, and $d_1(d_2)$ is the thickness of the top (bottom) layer BN flake.

The electric field in Fig. 2 is defined as the electric field in the TMDC heterostructure, which is given by $\frac{\varepsilon_{BN}}{2\varepsilon_{TMDC}}(\frac{V_{TG}}{d_1} - \frac{V_{BG}}{d_2})$.

### Data availability

Source data are available for this paper. The data in Figs.1-4 are provided in the source data files. All other data that support the plots within this paper and other findings of this study are available from the corresponding author upon reasonable request.

## Acknowledgments

We thank Prof. Chenhao Jin for the helpful discussions. Z.L. and S.-F.S. acknowledge support from NYSTAR through Focus Center-NY–RPI Contract C150117. The device fabrication was supported by the Micro and Nanofabrication Clean Room (MNCR) at Rensselaer Polytechnic Institute (RPI). S.-F.S. also acknowledges the support from NSF Grant DMR−1945420, DMR−2104902, and ECCS-2139692. X.H. and Y.-T.C. acknowledge support from NSF under awards DMR-2104805 and DMR-2145735. The optical spectroscopy measurements were supported by DURIP awards through Grant FA9550-20−1-0179 and FA9550-23-1-0084. S.T. acknowledges support from NSF DMR-1904716, DMR-1838443, CMMI-1933214, and DOE-SC0020653. K.W. and T.T. acknowledge support from JSPS KAKENHI (Grant Numbers 19H05790, 20H00354, and 21H05233). Y.S. and C.Z. acknowledge support from NSF PHY−2110212, PHY−1806227, OMR-2228725, ARO (W911NF17-1-0128), and AFOSR (FA9550−20−1-0220).

## Author contributions

S.-F.S. and Z. L. conceived the project. ZL, DC, and YM fabricated devices. ZL, DC, LM, LY, XH, and QW performed measurements. MB and ST grew the TMDC crystals. TT and KW grew the BN crystals. S.-F. S, Y.-T.C, ZL, DC, XC, Y. S., and C. Z. analyzed the data. S.-F. S and Y.-T. C. supervised the project. S.-F. S. and Y.-T. C. wrote the manuscript with inputs from all authors.

## Competing interests

The authors declare no competing interests.

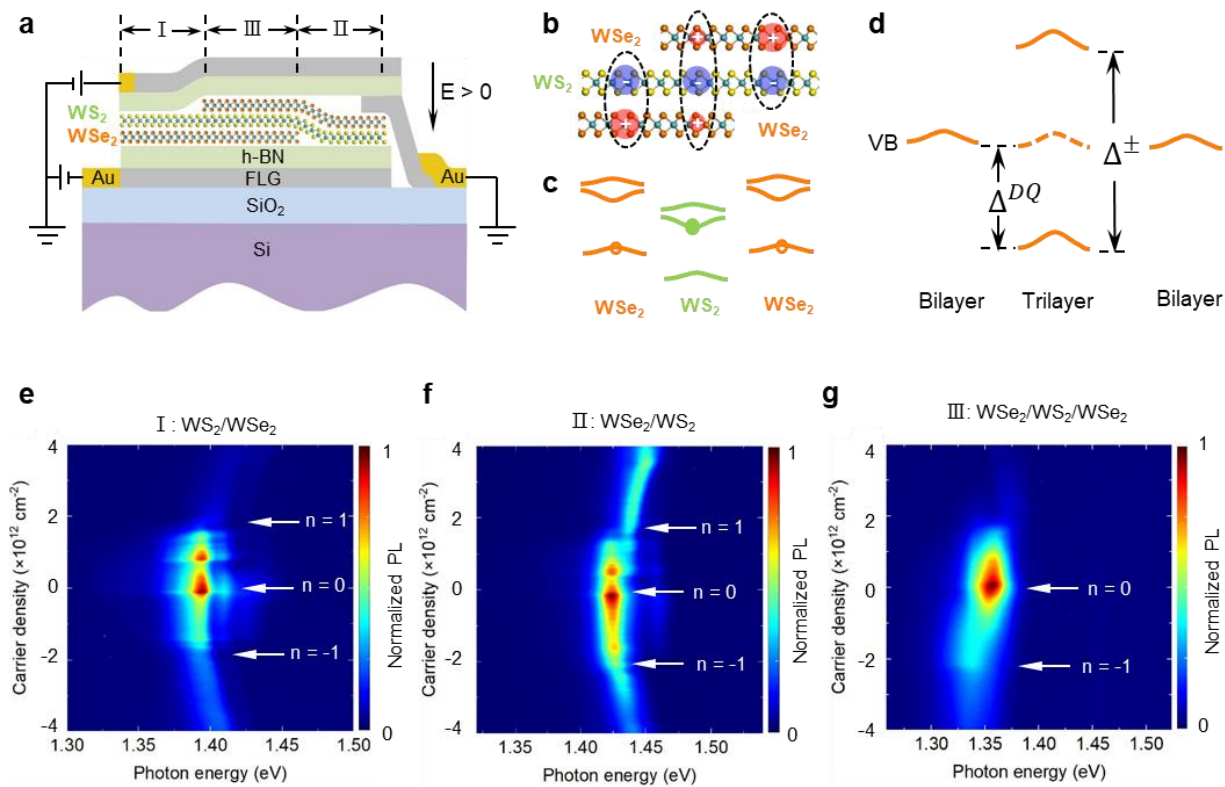

**Figure1. Excitons in different stacking structures of the trilayer device.** (a) Schematics of the device structure with three different regions: (I) $WS_2$/b-$WSe_2$ (II) t-$WSe_2$/$WS_2$/b-$WSe_2$ (III) top bilayer, t-$WSe_2$/$WS_2$. (b) Schematics of the dipolar and quadrupolar excitons configuration. (c) Type-II alignment of the angle-aligned $WSe_2$/$WS_2$ heterobilayer. (d) Valence band hybridization in the trilayer region, compared with the flat valence band of $WSe_2$ in the $WSe_2$/$WS_2$ moiré bilayer regions. (e), (f), and (g) are doping-dependent PL spectra for regions I, II, and III. The PL data were taken from device D5.

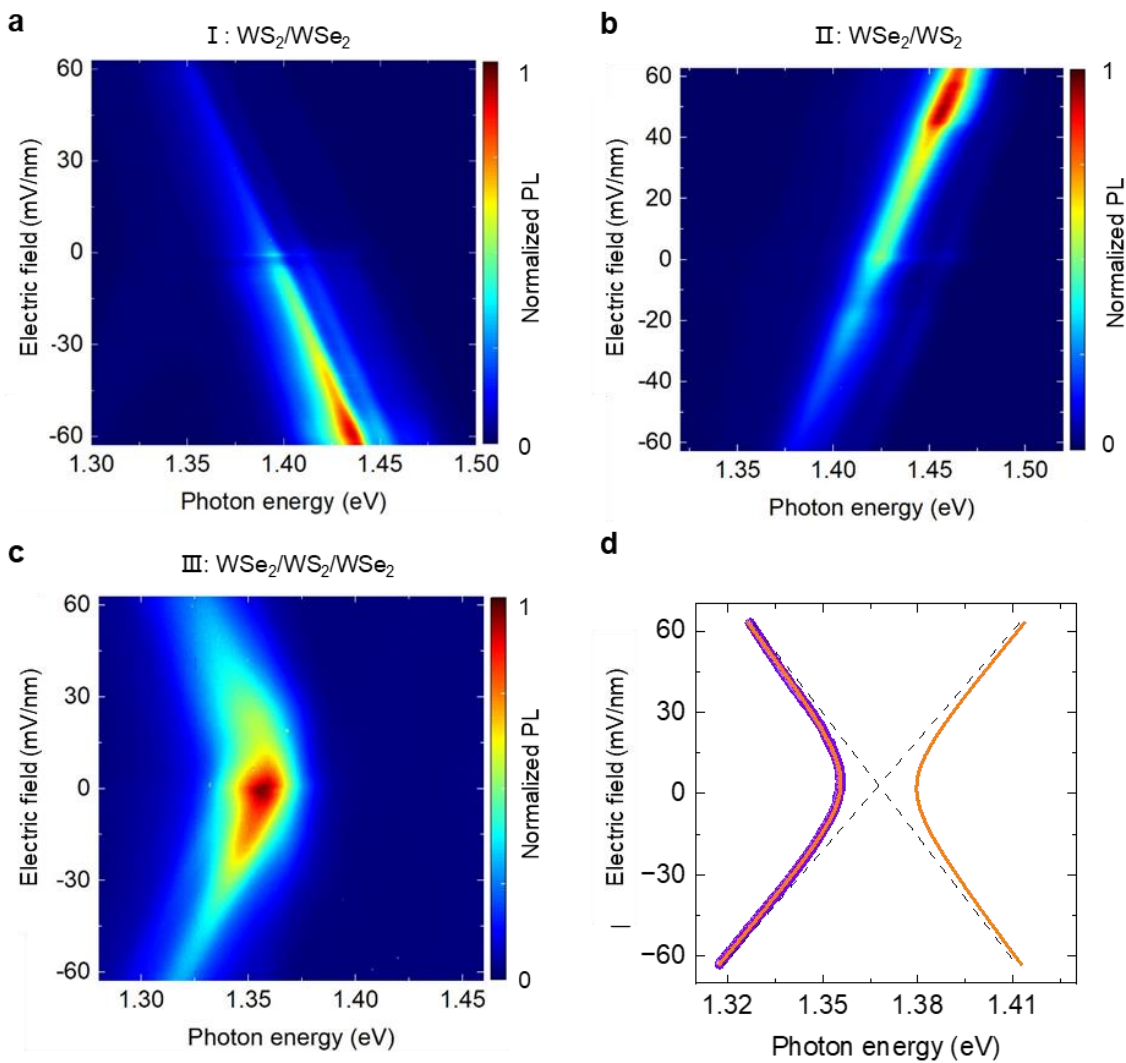

**Figure 2. Electric-field-dependent PL spectra of dipolar and quadrupolar excitons.**
(a), (b), and (c) are electric -field-dependent PL spectra of the region I, II, and III, respectively. (d) Fitting of the quadrupolar excitons PL resonances (orange curve) on extracted PL peak energy (purple spheres) as a function of the electric field from (c). The PL peak positions are extracted by fitting each PL spectrum with a single Lorentzian peak. The PL data were taken from device D5.

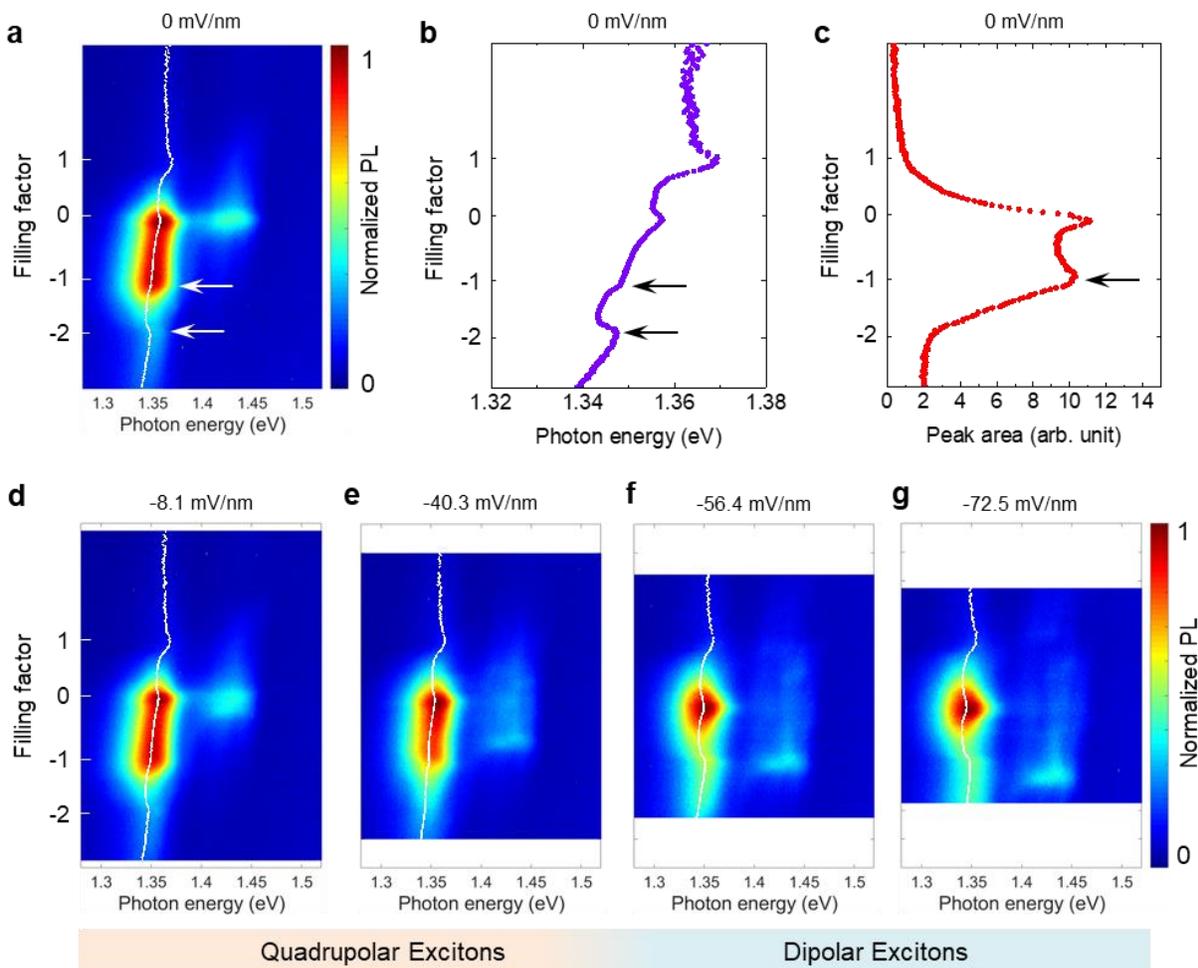

**Figure 3. Evolution of doping-dependent PL spectra of the trilayer region at different electric fields.** (a) the PL spectra as a function of the filling factor at the zero electric field. (b) PL peak energy and (c) integrated PL intensity, extracted from (a), plotted as a function of the filling factor. (d)-(g) are doping-dependent PL spectra at several external electric fields increasing in the negative direction. (d) and (e) are in the quadrupolar exciton regime, while (f) and (g) are transitioned to the dipolar exciton regime. The dotted white lines in the color plots are the extracted PL peak energies through fitting (details in Supplementary Information Section 8). The PL data were taken from device D1.

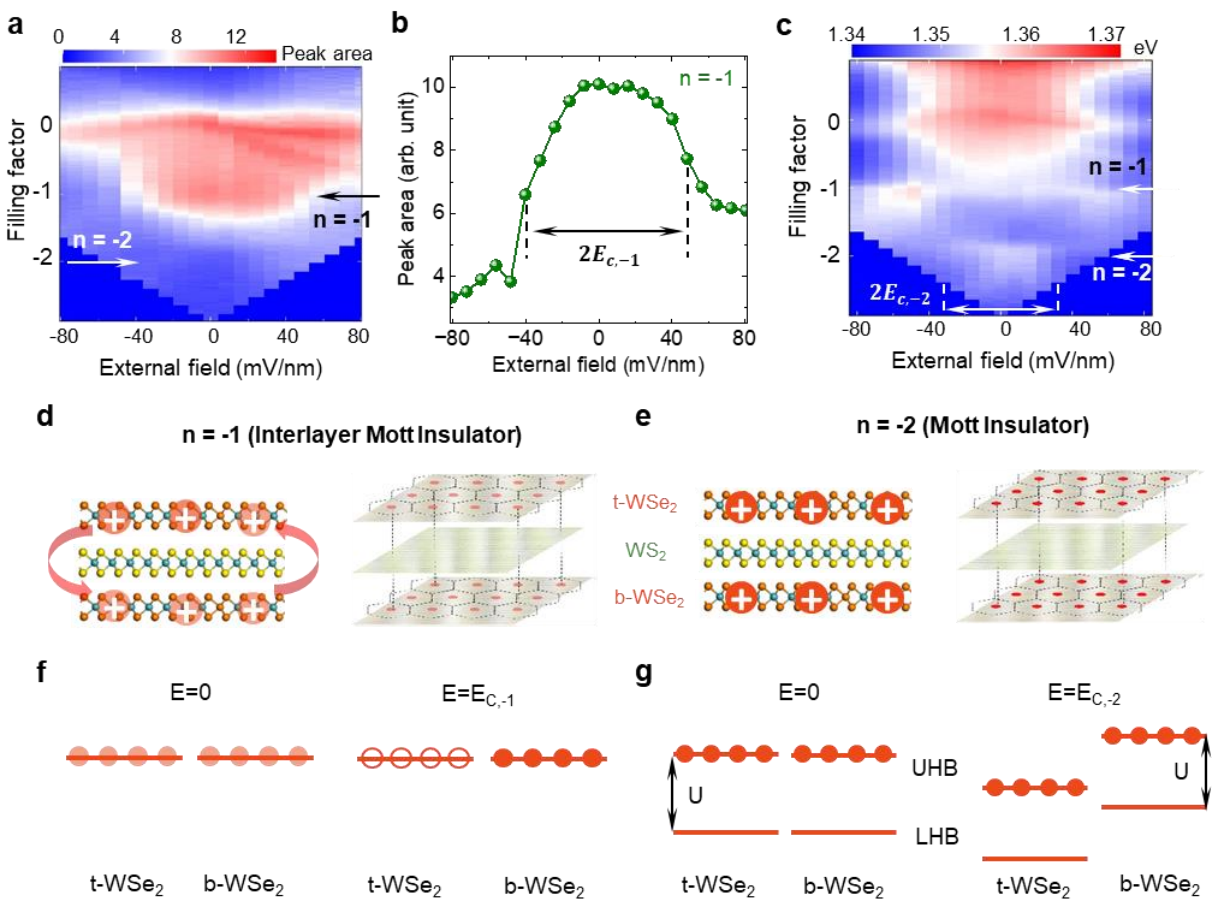

**Figure 4. Interlayer Mott insulator at n=-1 and Mott insulator at n=-2.** (a) The color plot of integrated PL intensity as a function of filling factor and external electric field. (b) The linecut of (a) at n=-1. (c) The color plot of PL peak energy as a function of the filling factor and external electric field. (d) and (e) are schematics of the hole configuration for the interlayer Mott insulator at n=-1 and Mott insulator at n=-2, respectively. (f) and (g), are the evolution of alignment and filling of the flat valence bands of the t-WSe₂ and b-WSe₂ layers as the electric field increases for n=-1 (f) and n=-2 (g). The PL data were taken from device D1.

## References

1. Xu, X., Yao, W., Xiao, D. & Heinz, T. F. Spin and pseudospins in layered transition metal dichalcogenides. *Nat Phys* **10**, 343–350 (2014).

2. Mak, K. F. & Shan, J. Photonics and optoelectronics of 2D semiconductor transition metal dichalcogenides. *Nat. Photonics* **10**, 216–226 (2016).

3. Lee, C. *et al.* Atomically thin p − n junctions with van der Waals heterointerfaces. *Nat. Nanotechnol.* **9**, 676–681 (2014).

4. Hong, X. *et al.* Ultrafast charge transfer in atomically thin $MoS_2$/$WS_2$ heterostructures. *Nat. Nanotechnol.* **9**, 682 (2014).

5. Rivera, P. *et al.* Observation of long-lived interlayer excitons in monolayer MoSe2-WSe2 heterostructures. *Nat. Commun.* **6**, 6242 (2015).

6. Rivera, P. *et al.* Interlayer valley excitons in heterobilayers of transition metal dichalcogenides. *Nat. Nanotechnol.* **13**, 1004–1015 (2018).

7. Regan, E. C. *et al.* Mott and generalized Wigner crystal states in WSe2/WS2 moiré superlattices. *Nature* **579**, 359–363 (2020).

8. Tang, Y. *et al.* Simulation of Hubbard model physics in WSe2/WS2 moiré superlattices. *Nature* **579**, 353–358 (2020).

9. Campbell, A. J. *et al.* Exciton-polarons in the presence of strongly correlated electronic states in a MoSe2/WSe2 moiré superlattice. *npj 2D Mater. Appl.* **6**, 1–8 (2022).

10. Gu, J. *et al.* Dipolar excitonic insulator in a moiré lattice. *Nat. Phys.* **18**, 395–400 (2022).

11. Zhang, Z. *et al.* Correlated interlayer exciton insulator in heterostructures of monolayer WSe2 and moiré $WS_2$/$WSe_2$. *Nat. Phys.* **18**, 1214–1220 (2022).

12. Huang, X. *et al.* Correlated Insulating States at Fractional Fillings of the WS2/WSe2 Moiré Lattice. *Nat. Phys.* **17**, 715–719 (2021).

13. Zhou, Y. *et al.* Bilayer Wigner crystals in a transition metal dichalcogenide heterostructure. *Nature* **595**, 48–52 (2021).

14. Smoleński, T. *et al.* Signatures of Wigner crystal of electrons in a monolayer semiconductor. *Nature* **595**, 53–57 (2021).

15. Wang, L. *et al.* Correlated electronic phases in twisted bilayer transition metal dichalcogenides. *Nat. Mater.* **19**, 861–866 (2020).

16. Ghiotto, A. *et al.* Quantum criticality in twisted transition metal dichalcogenides. *Nature* **597**, 345–349 (2021).

17. Xu, Y. *et al.* Correlated insulating states at fractional fillings of moiré superlattices. *Nature* **587**, 214–218 (2020).

18. Yuan, H. *et al.* Zeeman-type spin splitting controlled by an electric field. *Nat. Phys.* **9**, 563–569 (2013).

19. Zeng, Y. *et al.* Exciton density waves in Coulomb-coupled dual moiré lattices. *Nat. Mater.* **22**, 175–179 (2023).

20. Rivera, P. *et al.* Valley-Polarized Exciton Dynamics in a 2D Semicondcutor Heterostructure. *Science* **351**, 688 (2016).

21. Jauregui, L. A. *et al.* Electrical control of interlayer exciton dynamics in atomically thin heterostructures. *Science* **366**, 870–875 (2019).

22. Seyler, K. L. *et al.* Signatures of moiré-trapped valley excitons in MoSe 2 /WSe 2 heterobilayers. *Nature* **567**, 66–70 (2019).

23. Tran, K. *et al.* Evidence for moiré excitons in van der Waals heterostructures. *Nature* **567**, 71–75 (2019).

24. Wu, F., Lovorn, T. & MacDonald, A. H. Topological Exciton Bands in Moiré Heterojunctions. *Phys. Rev. Lett.* **118**, 147401 (2017).

25. Eisenstein, J. P. & Macdonald, A. H. Bose–Einstein condensation of excitons in bilayer electron systems. *Nautre* **432**, 691–694 (2004).

26. Wang, Z. *et al.* Evidence of high-temperature exciton condensation in two-dimensional atomic  double layers. *Nature* **574**, 76–80 (2019).

27. Fogler, M. M., Butov, L. V. & Novoselov, K. S. High-temperature superfluidity with indirect excitons in van der Waals heterostructures. *Nat. Commun.* **5**, Article number: 4555 (2014).

28. Zhang, Y. H., Sheng, D. N. & Vishwanath, A. SU(4) Chiral Spin Liquid, Exciton Supersolid, and Electric Detection in Moiré Bilayers. *Phys. Rev. Lett.* **127**, 247701 (2021).

29. Slobodkin, Y. *et al.* Quantum Phase Transitions of Trilayer Excitons in Atomically Thin Heterostructures. *Phys. Rev. Lett.* **125**, 255301 (2020).

30. Astrakharchik, G. E., Kurbakov, I. L., Sychev, D. V., Fedorov, A. K. & Lozovik, Y. E. Quantum phase transition of a two-dimensional quadrupolar system. *Phys. Rev. B* **103**, L140101 (2021).

31. Miao, S. *et al.* Strong interaction between interlayer excitons and correlated electrons in WSe 2 /WS 2 moiré superlattice. *Nat. Commun.* **12**, Article number: 3608 (2021).

32. Liu, E. *et al.* Excitonic and Valley-Polarization Signatures of Fractional Correlated Electronic Phases in a WSe2/WS2 Moiré Superlattice. *Phys. Rev. Lett.* **127**, 37402 (2021).

33. Wang, X. *et al.* Intercell moiré exciton complexes in electron lattices. *Nat. Mater.* **22**, 599 (2023).

34. Baek, H. *et al.* Optical read-out of Coulomb staircases in a moiré superlattice via trapped interlayer trions. *Nat. Nanotechnol.* **16**, 1237–1243 (2021).

35. Ciarrocchi, A. *et al.* Polarization switching and electrical control of interlayer excitons in two-dimensional van der Waals heterostructures. *Nat. Photonics* **13**, 131–136 (2019).

36. Tang, Y. *et al.* Tuning layer-hybridized moiré excitons by the quantum-confined Stark effect. *Nat. Nanotechnol.* **16**, 52–57 (2021).

37. Kim, B., Watanabe, K., Taniguchi, T., Barmak, K. & Lui, C. H. Optical absorption of interlayer excitons in transition-metal dichalcogenide heterostructures. *Science* **410**, 406–410 (2022).

38. Li, W., Lu, X., Dubey, S., Devenica, L. & Srivastava, A. Dipolar interactions between localized interlayer excitons in van der Waals heterostructures. *Nat. Mater.* **19**, 624–629 (2020).

39. Leisgang, N. *et al.* Giant Stark splitting of an exciton in bilayer MoS2. *Nat. Nanotechnol.* **15**, 901–907 (2020).

40. Wang, T. *et al.* Giant Valley-Zeeman Splitting from Spin-Singlet and Spin-Triplet Interlayer Excitons in WSe2/MoSe2 Heterostructure. *Nano Lett.* **20**, 694–700 (2020).

41. Chen, D. *et al.* Tuning moiré excitons and correlated electronic states through layer degree of freedom. *Nat. Commun.* **13**, Article number: 4810 (2022).

42. Chen, D. *et al.* Excitonic insulator in a heterojunction moiré superlattice. *Nat. Phys.* **18**, 1171–1176 (2022).

43. Wu, F., Lovorn, T., Tutuc, E., Martin, I. & Macdonald, A. H. Topological Insulators in Twisted Transition Metal Dichalcogenide Homobilayers. *Phys. Rev. Lett.* **122**, 86402 (2019).

44. Xiong, R. *et al.* Correlated insulator of excitons in WSe$_2$/WS$_2$ moiré superlattices. *Science* **380**, 860–864 (2023).

45. Li, X., Wu, F. & MacDonald, A. H. Electronic structure of single-twist trilayer graphene. arXiv:1907.12338 (2019).

46. Bai, Y. *et al.* Evidence for exciton crystals in a 2D semiconductor heterotrilayer. *Arxiv* arXiv:2207.09601 (2022) doi:arXiv:2207.09601.

47. Li, W. *et al.* Quadrupolar excitons in a tunnel-coupled van der Waals heterotrilayer. *arxiv* ArXiv:2208.05490 (2022).

## Supplementary Information

## Quadrupolar Excitons and Hybridized Interlayer Mott Insulator in a Trilayer Moiré Superlattice

Zhen Lian[1#], Dongxue Chen[1#], Lei Ma[1#], Yuze Meng[1#], Ying Su[2], Li Yan[1], Xiong Huang[3,4], Qiran Wu[3], Xinyue Chen[1], Mark Blei[5], Takashi Taniguchi[6], Kenji Watanabe[7], Sefaattin Tongay[5], Chuanwei Zhang[2], Yong-Tao Cui[3*], Su-Fei Shi[1,8*]

1. Department of Chemical and Biological Engineering, Rensselaer Polytechnic Institute, Troy, NY 12180, USA
2. Department of Physics, University of Texas at Dallas, Dallas, Texas, 75083, USA
3. Department of Physics and Astronomy, University of California, Riverside, California, 92521, USA
4. Department of Materials Science and Engineering, University of California, Riverside, California, 92521, USA
5. School for Engineering of Matter, Transport and Energy, Arizona State University, Tempe, AZ 85287, USA
6. International Center for Materials Nanoarchitectonics, National Institute for Materials Science, 1-1 Namiki, Tsukuba 305-0044, Japan
7. Research Center for Functional Materials, National Institute for Materials Science, 1-1 Namiki, Tsukuba 305-0044, Japan
8. Department of Electrical, Computer & Systems Engineering, Rensselaer Polytechnic Institute, Troy, NY 12180, USA

[#] These authors contributed equally to this work
[*] Corresponding authors: yongtao.cui@ucr.edu, shis2@rpi.edu

## Supplementary section 1: Optical microscope image of device D5 and D1

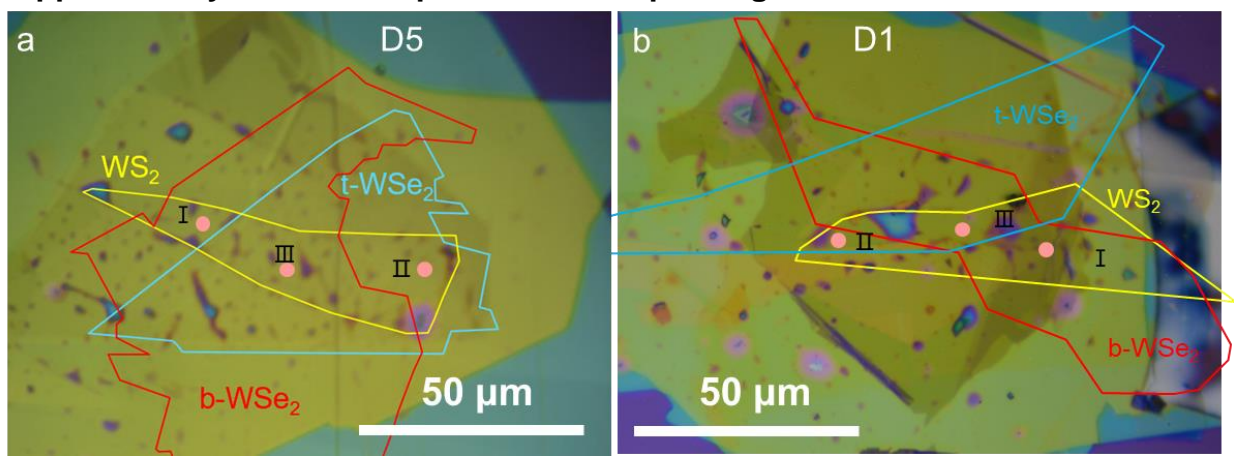

**Figure S1. Optical microscope images of the devices shown in the main text.** (a) and (b) are the optical images of device D5 and D1, respectively. The pink dots indicate the laser excitation spots for optical measurements.

## Supplementary section 2: SHG measurement and determination of alignment angles

Figure S2a and c show normalized SHG intensities as functions of the half-waveplate angle measured on different regions of device D5 and D1. The orientation of the armchair direction of each layer can be obtained by fitting the data with a sinusoidal function.

For device D5, we obtained 1.4° ± 0.5° alignment between b-WSe$_2$ and WS$_2$ and -0.1°± 0.5° alignment between t-WSe$_2$ and WS$_2$.

For device D1, we obtained 0.9° ± 0.5° alignment between b-WSe$_2$ and region I and 1.2° ± 0.5° alignment between t-WSe$_2$ and region I.

The SHG intensity from region I and region II (t-WSe$_2$/WS$_2$) are enhanced compared to the SHG intensities from both WSe$_2$ layers, indicating a close to 0° alignment for both heterobilayers (Figure S2b and d) in D5 and D1.

The lattice constant of WSe$_2$ and WS$_2$, $a_{Se}$ and $a_S$ were chosen to 0.328 nm and 0.315 nm. The moiré lattice constant can be calculated by $a_m = \frac{4\pi}{\sqrt{3}|\mathbf{k}_{Se}-\mathbf{k}_S|}$, where $\mathbf{k}_{Se}$ and $\mathbf{k}_S$ are reciprocal primitive vectors of WSe$_2$ and WS$_2$, respectively. The moiré density $n_0$ is given by $\frac{2}{\sqrt{3}a_m^2}$ . Using the alignment angles determined by SHG, we estimate $n_0 = 2.1 \times 10^{12} cm^{-2}$ for region I and $n_0 = 2.3 \times 10^{12} cm^{-2}$ for region II for device D1 shown in Figs.3 and 4.

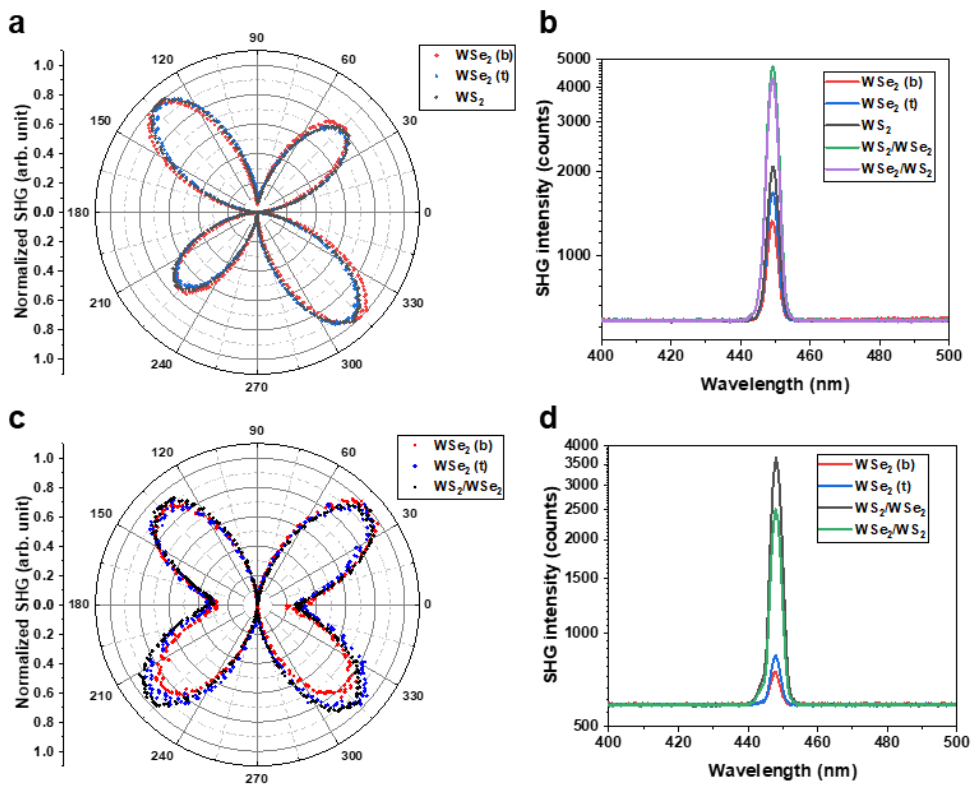

**Figure S2. Polarized SHG measurement from different regions of device D5 and D1 shown in main text.** (a) shows the SHG intensity as a function of half-waveplate angle measured on b-WSe$_2$, t-WSe$_2$ and WS$_2$ of device D5. (b) shows the relative SHG intensity from b-WSe$_2$, t-WSe$_2$, WS$_2$, region I (WS$_2$/ WSe$_2$) and region II (WSe$_2$/ WS$_2$) of device D5. (c) shows the SHG intensity as a function of half-waveplate angle measured on b-WSe$_2$, t-WSe$_2$ and region I (WS$_2$/WSe$_2$) of device D1. (d) shows the relative SHG intensity from b-WSe$_2$, t-WSe$_2$, region I (WS$_2$/ WSe$_2$) and region II (WSe$_2$/ WS$_2$) of device D1. The enhancement of SHG signal on region I and region II indicates the top two layers and the bottom two layers are both close-to-zero-degree-aligned in both devices.

**Supplementary section 3: Comparison of the PL spectra at zero gate voltage in the bilayer regions and the trilayer region**

Figure S3 shows the PL spectra from region I (WS$_2$/b-WSe$_2$), region II (t-WSe$_2$/WS$_2$) and region III (t-WSe$_2$/WS$_2$/b-WSe$_2$) of device D5 at zero gate voltage. All the PL spectra were taken at 6 K using 50 uW 633 nm excitation and the same integration time (2s).

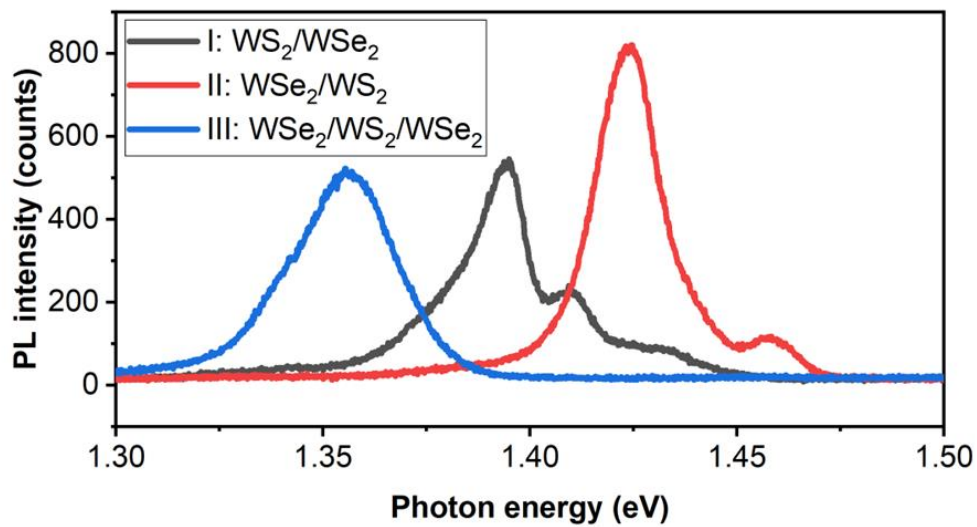

**Figure S3**. Comparison of the PL spectra from region I, region II and region III at zero gate voltages.

## Supplementary section 4: PL spectra from different regions of device D1

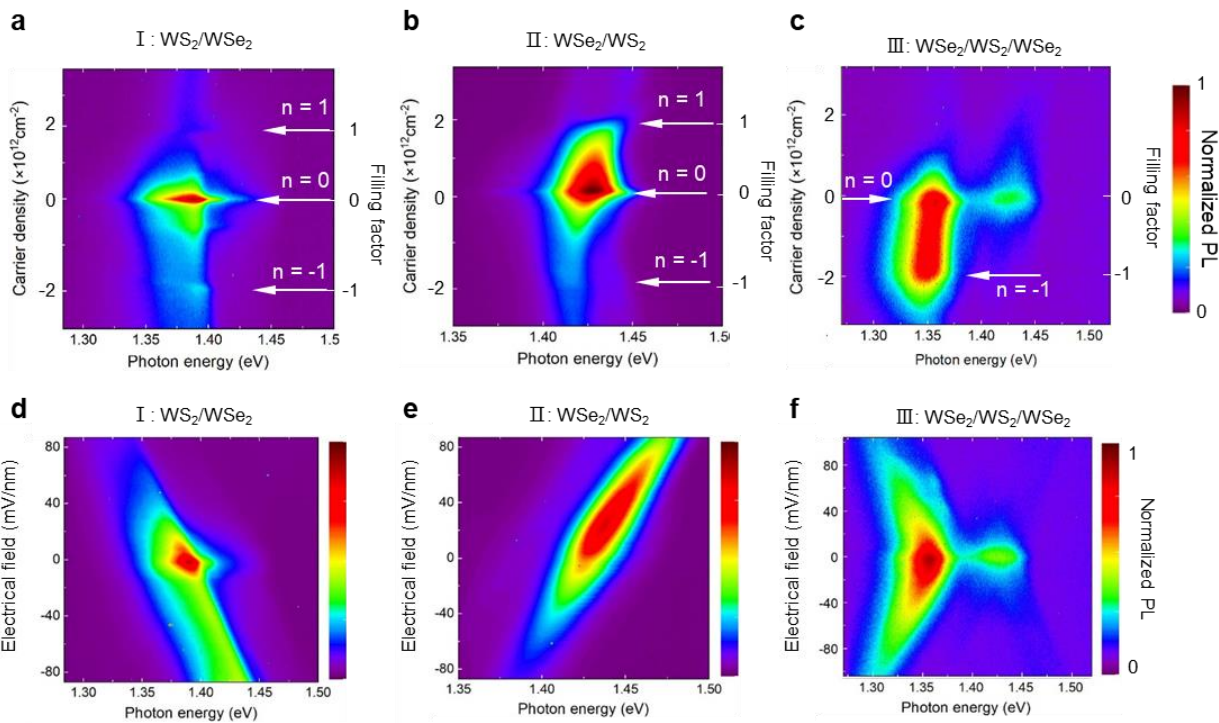

**Figure S4. Doping-dependent and electric-field-dependent PL spectra from device D1.** (a), (b) and (c) show the doping-dependent PL spectra taken from region I, II and III of device D1, respectively. (d), (e) and (f) show the electric-field-dependent PL spectra taken from region I, II and III of device D1, respectively.

## Supplementary section 5: Power dependence of the PL spectra from device D1

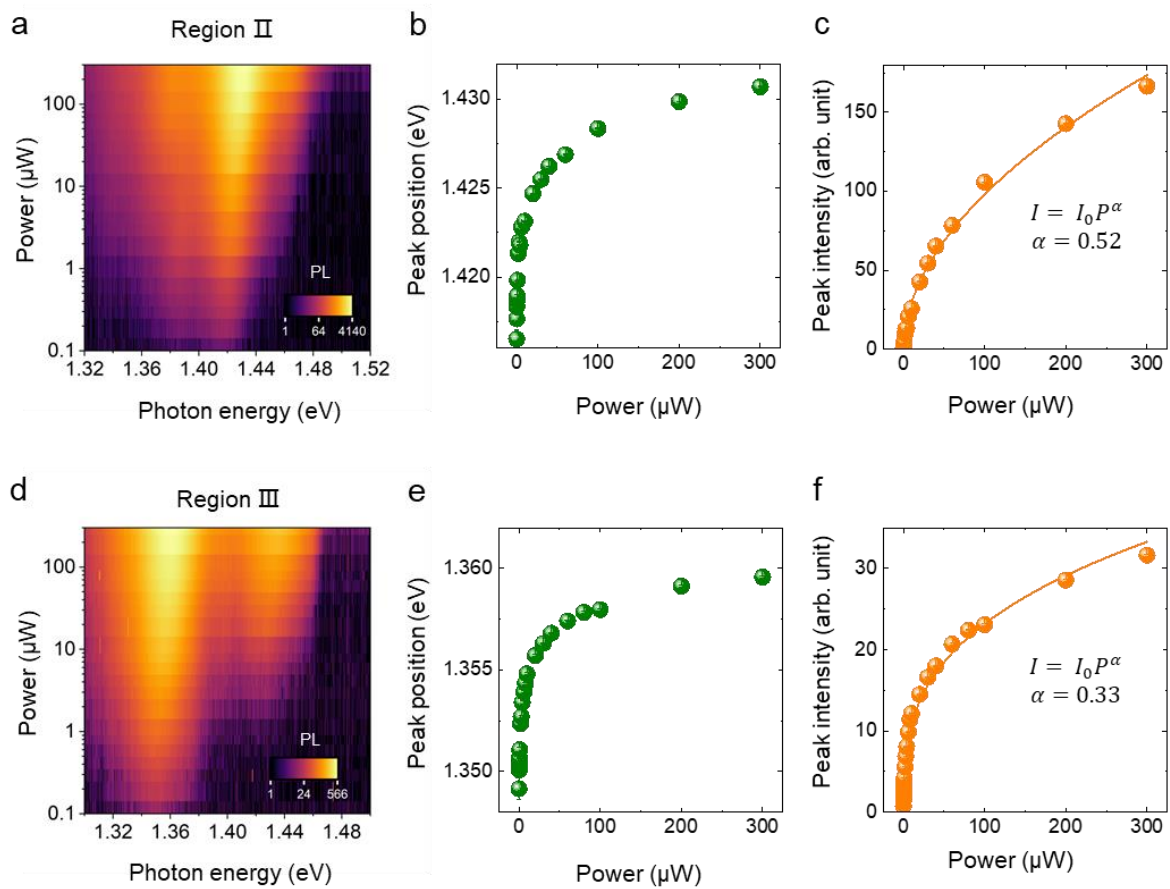

**Figure S5.** Excitation power dependence of PL from region II and region III (Device D1, shown in the main text). (a) and (d) are PL spectra for different excitation powers from region II and III, respectively. (b) and (e) are PL peak position as a function of excitation power from region II and III. (c) and (f) are integrated PL intensity of excitons as a function of the excitation power from region II and III. The data are taken with the CW excitation with phonon energy centered at 1.959 eV and temperature of 6 K.

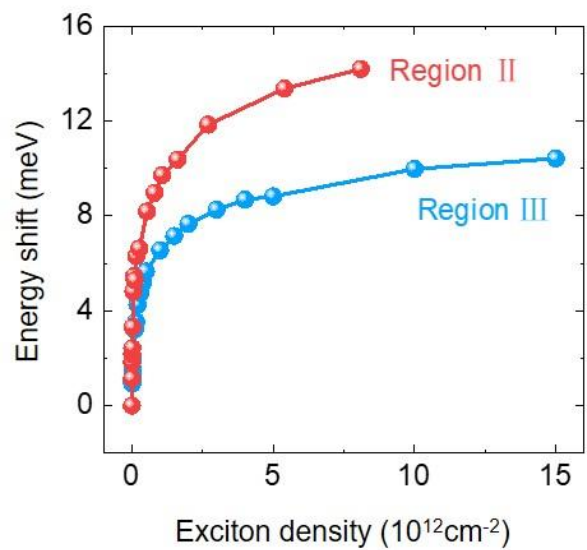

**Figure S6.** PL peak blueshifts as a function of the exciton density for device D1.

## Supplementary section 6: Temperature dependence of the PL spectra from device D1

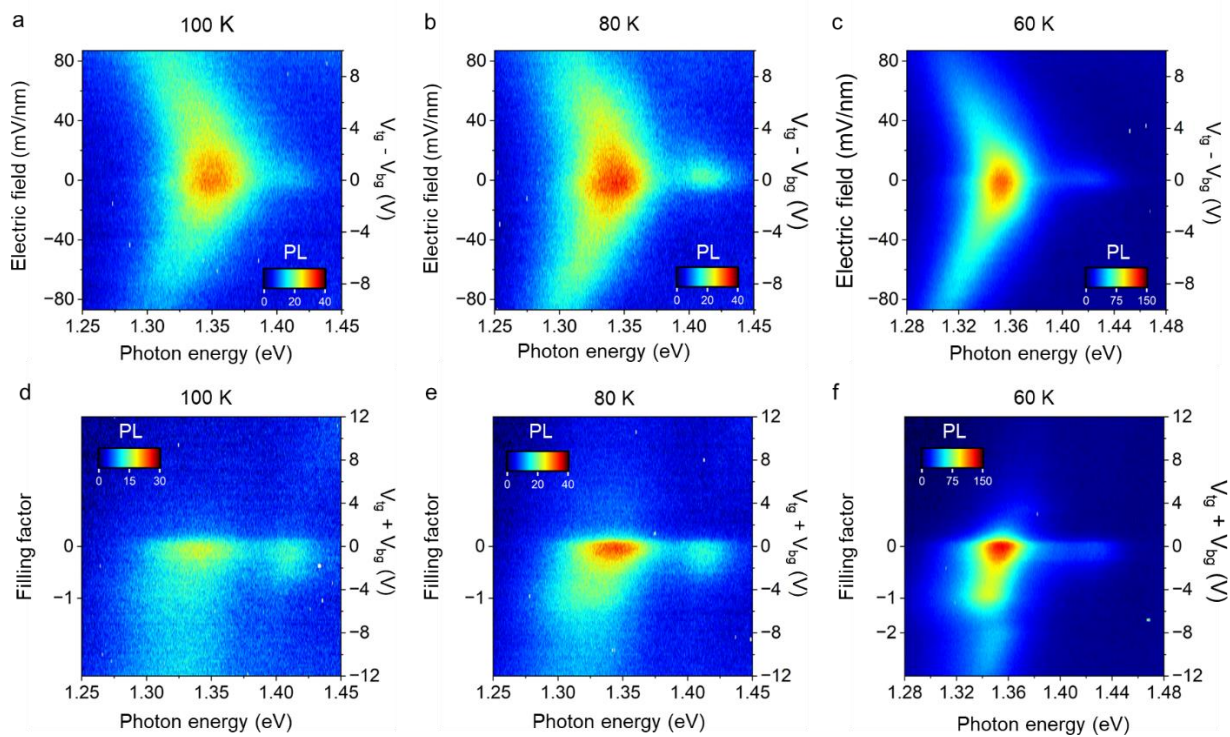

**Figure S7.** PL spectra of region III measured at different temperatures for device D1. (a)-(c) are the electric-field-dependent PL spectra. (d)-(f) are the doping-dependent PL spectra.

## Supplementary section 7: Reflectance contrast measurement and determination of filling factors for device D1

Top-gate-dependent, back-gate-dependent and dual-gate-dependent reflectance contrast spectra with $V_{tg} = V_{bg}$ measured on region I ($WS_2/WSe_2$) are shown in Figure S8a,b,c. The gate voltages corresponding to n=-1, n=+1 and n=0 can be identified by the abrupt change of resonance positions and intensity in the reflectance contrast spectra, consistent with previous report[1,2]. The gate voltages corresponding to different filling factors are the same in the top-gate-only and bottom-gate-only configuration, indicating an equal thickness of the top BN and the bottom BN. The gate dependence in Figure S8 indicates that the gate voltage between the filling -2, -1, 0 and +1 is ~ 3 V. Using the geometry capacitance model, we estimate the thickness of BN layer to be ~ 31 nm.

The filling factor as a function of gate voltages in Figure S8c is used to calibrate the filling factor of the PL spectra measured from region III, which is done using the same gate configuration. It can be found that the peak position and intensity change in the PL spectra is aligned with the filling factors calibrated with this method.

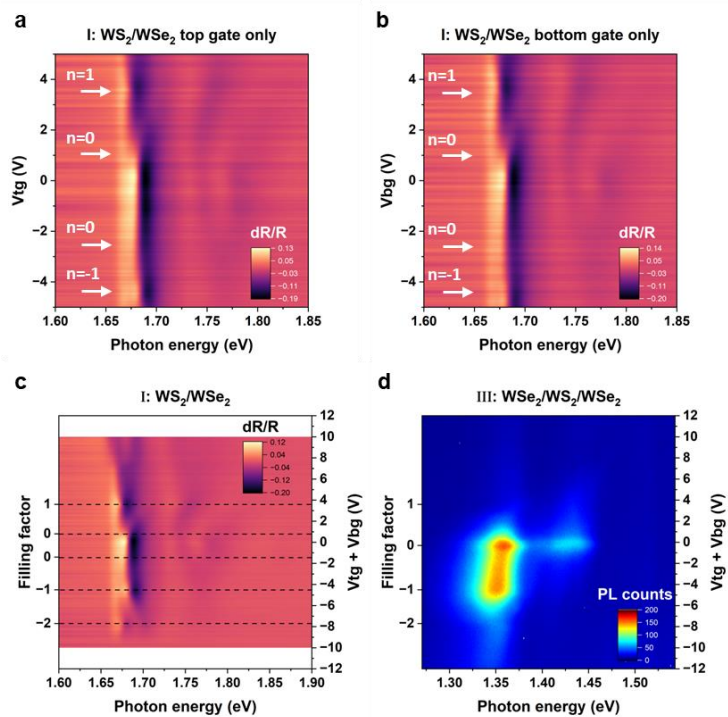

**Figure S8. Reflection contrast measurement from region I of device D1 and calibration of filling factors.** (a) and (b) show gate-dependent reflection contrast spectra in the region I measured with the top gate and back gate, respectively. (c) shows the reflectance contrast spectra as a function of $V_{tg}+V_{bg}$ while keeping $V_{tg}$ and $V_{bg}$ equal. The dashed lines indicate the filling factors determined from the spectra. (d) shows the gate-dependent PL spectra of device D1. It can be found that at the integer fillings determined in (c), the PL position and intensity exhibit abrupt changes.

## Supplementary section 8: Doping-dependent PL spectra at different electric fields and fitting results

Figure S9 shows the PL spectra as functions of the filling factor in the trilayer region (region III) at different positive electric fields. Figure S10 shows the PL spectra as functions of the filling factor from the trilayer region (region III) at different negative electric fields. The PL from the lower energy branch can be fitted with a single Lorentzian peak. The fitted PL peak positions and integrated PL intensities for different electric fields are displayed in Figure S11 and Figure S12, respectively.

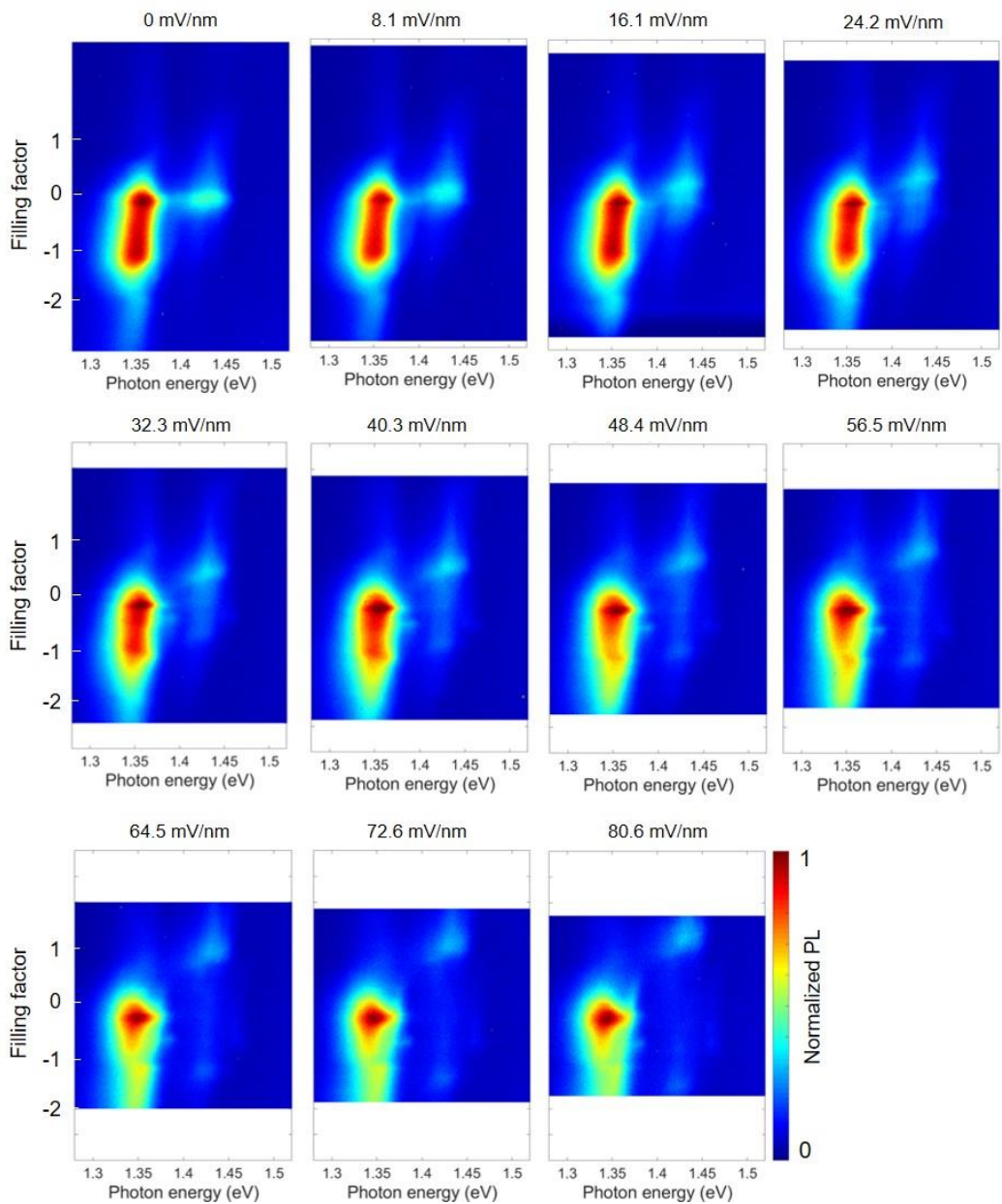

**Figure S9.** PL spectra as a function of the filling factor from the trilayer region at different positive electric fields.

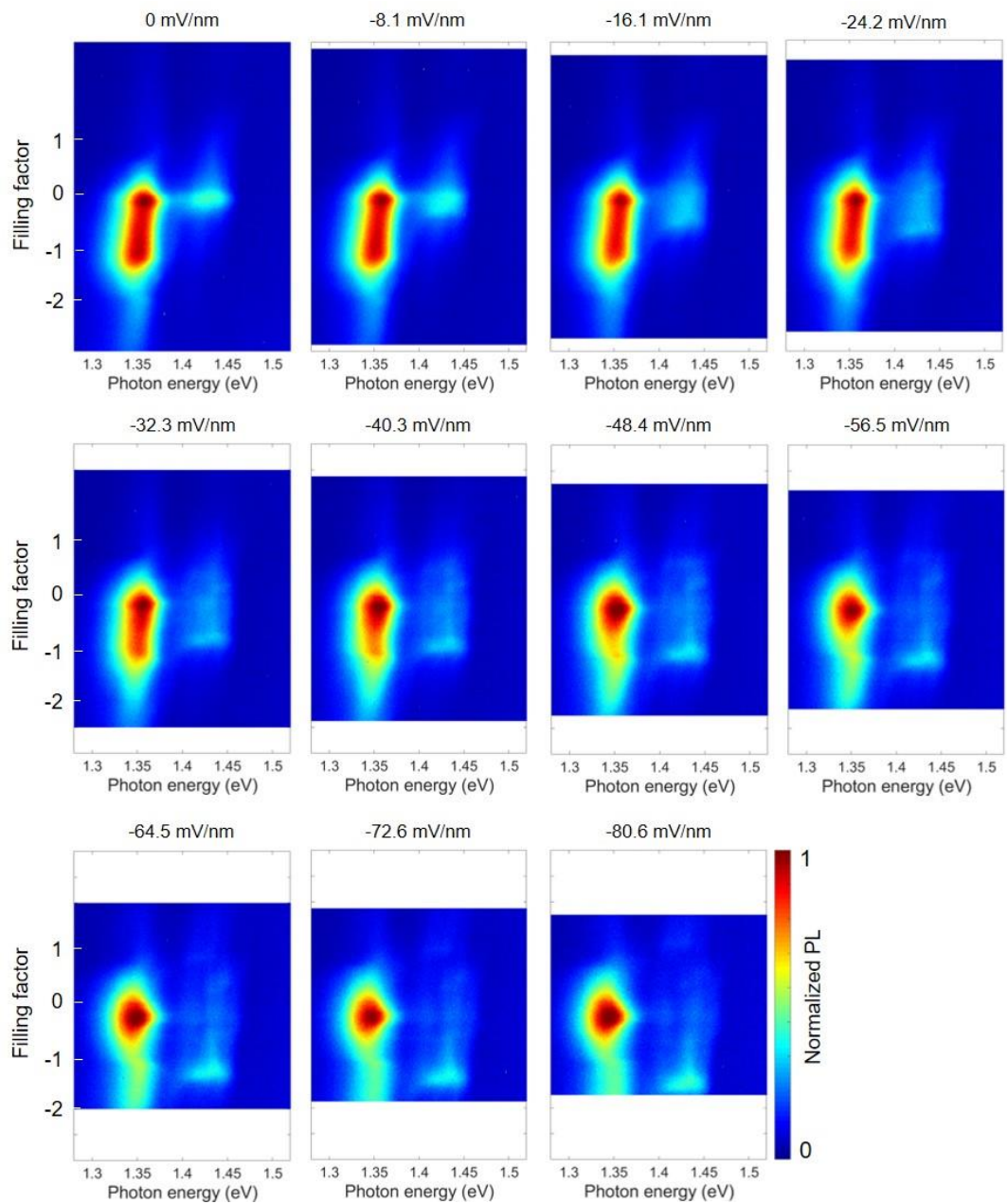

**Figure S10.** PL spectra as a function of the filling factor from the trilayer region at different negative electric fields.

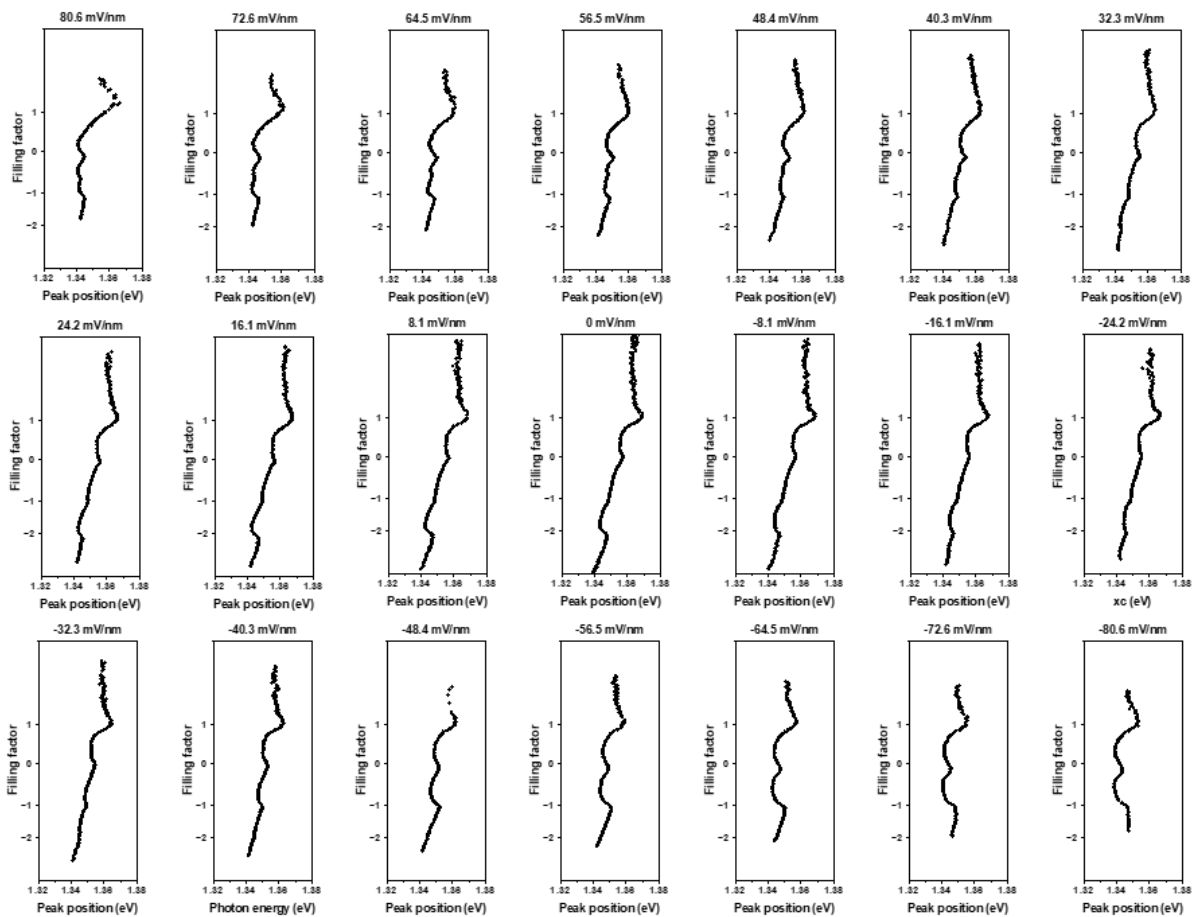

**Figure S11.** PL Peak positions extracted from Figure S9 and Figure S10 through fitting.

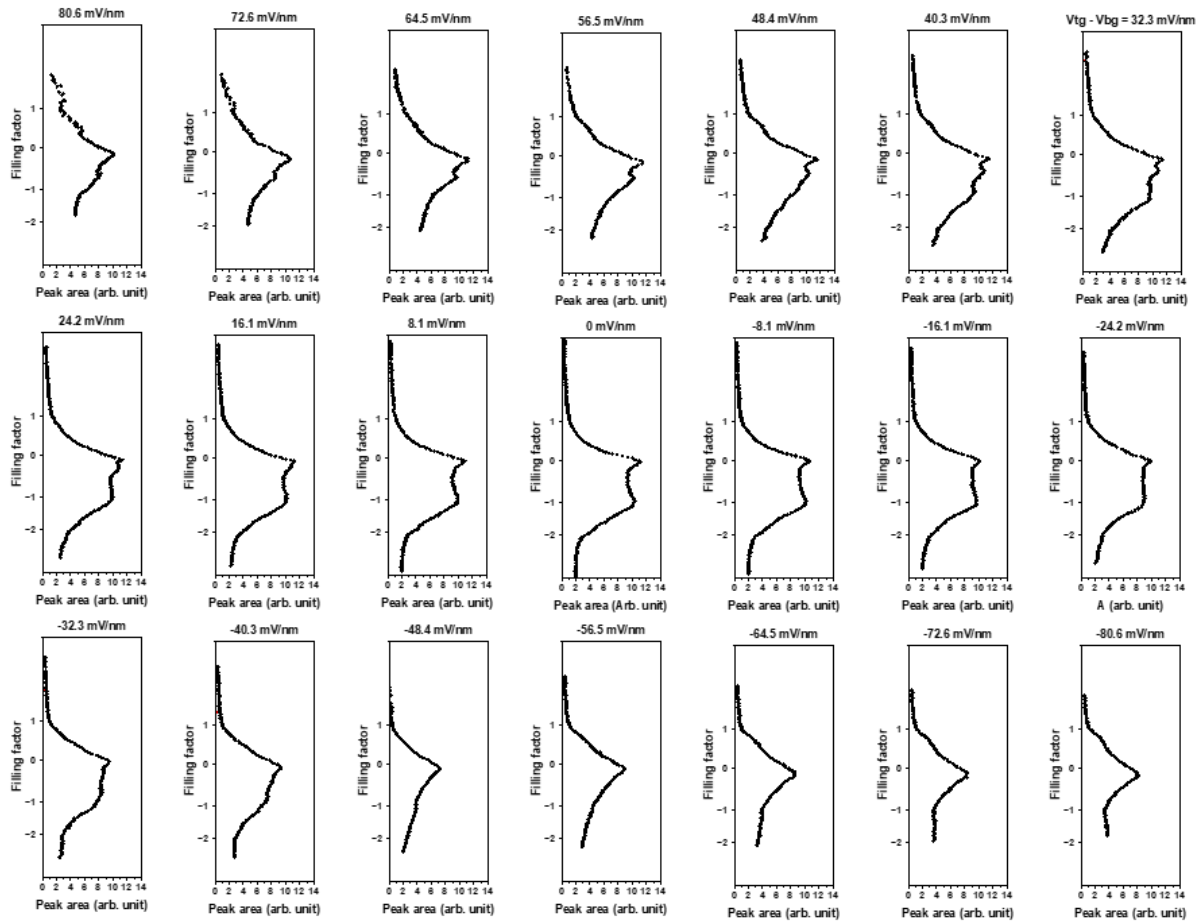

**Figure S12.** Integrated PL intensity extracted from Figure S9 and Figure S10 through the fitting.

## Supplementary section 9: Modeling of quadrupolar excitons

We consider the following two-energy-level Hamiltonian describing the coupling between the two dipolar interlayer excitons:

$$H = \begin{bmatrix} X_1(E) & \Delta \\ \Delta & X_2(E) \end{bmatrix} \tag{9.1}$$

Where $X_1(E)$ and $X_2(E)$ are the dipolar interlayer exciton energies at local electric field E in the WS$_2$/b-WSe$_2$ and t-WSe$_2$/WS$_2$ heterostructures, respectively, $\Delta$ is the coupling between the two interlayer excitons. Solving the eigenvalues of this system, we obtain the expression of the low-energy branch of the hybridized excitons:

$$X_{LE} = \frac{X_1(E) + X_2(E)}{2} - \frac{1}{2}\sqrt{(X_1(E) - X_2(E))^2 + 4\Delta^2} \tag{9.2}$$

The energies of the dipolar excitons can be expressed as:

$$X_1(E) = X_1 - Eed_1 \tag{9.3}$$

$$X_2(E) = X_2 + Eed_2 \tag{9.4}$$

Where $X_1$ and $X_2$ are the two dipolar exciton energies at zero electric field.

Eqn. 9.2. is fitted to the extracted peak positions at different electric fields shown in Fig.2d, with $X_1$, $X_2$, $d_1$, $d_2$ and $\Delta$ as fitting parameters. For device D5 shown in Fig.2d, we obtain $X_1 = 1.3696 \pm 0.0002\ eV$ , $X_2 = 1.3654 \pm 0.0002\ eV$ , $d_1 = 0.659 \pm 0.003\ nm$ , $d_2 = 0.740 \pm 0.003\ nm$ and $\Delta = 12.0 \pm 0.2\ meV$.

## Supplementary section 10: Electric field dependent PL spectra in WSe$_2$/WS$_2$/WSe$_2$ device D2

Figure S13a shows the electric field dependent PL spectra in another WSe$_2$/WS$_2$/WSe$_2$ device. Figure S13b shows the fitting result of the two-level hybridization model. We obtain $\Delta = 30 \pm 9$ meV through the fitting shown in Figure S13b.

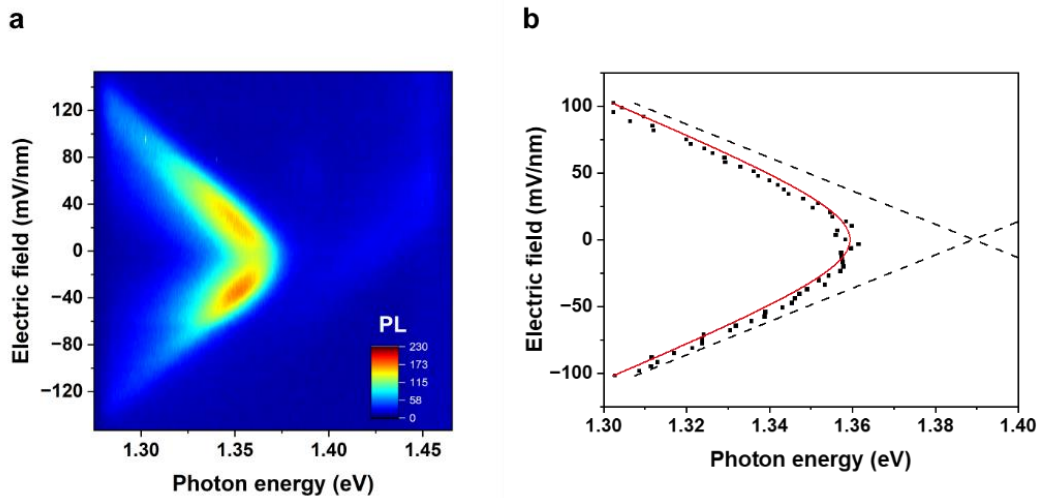

**Figure S13. Electric field dependent PL from quadrupolar excitons in another WSe₂/WS₂/WSe₂ device (D2).** (a) shows the electric field dependent PL spectra from the device D2. (b) shows the fitting result of the two-level hybridization model. The data are taken with a 50 µW CW excitation with photon energy centered at 1.959 eV and a temperature of 9 K.

## Supplementary section 11: Calculation of the critical electric field in the TMDC heterostructure at n=-1 and n=-2

We define $E_t$, $E_b$, and $E_h$ as the electric field inside top BN, bottom BN and TMDC heterostructure, $d_t$, $d_b$ and $d_h$ as the thickness of top BN, bottom BN and the TMDC heterostructure. Here in our case $d_t = d_b = d \approx 31$ nm. The dielectric constants of BN and TMDC are denoted as $\varepsilon_{BN}$ and $\varepsilon_{TMDC}$, respectively. Here we consider the electric field inside the TMDC heterostructure in three special cases: (1) the heterostructure is charge-neural; (2) n = -1 and all the holes reside in the b-WSe₂ layer; and (3) n=-2 and holes are distributed equally in the top and bottom WSe₂. The corresponding charge and electric field distribution are shown in Figure S14 (a), (b) and (c), respectively.

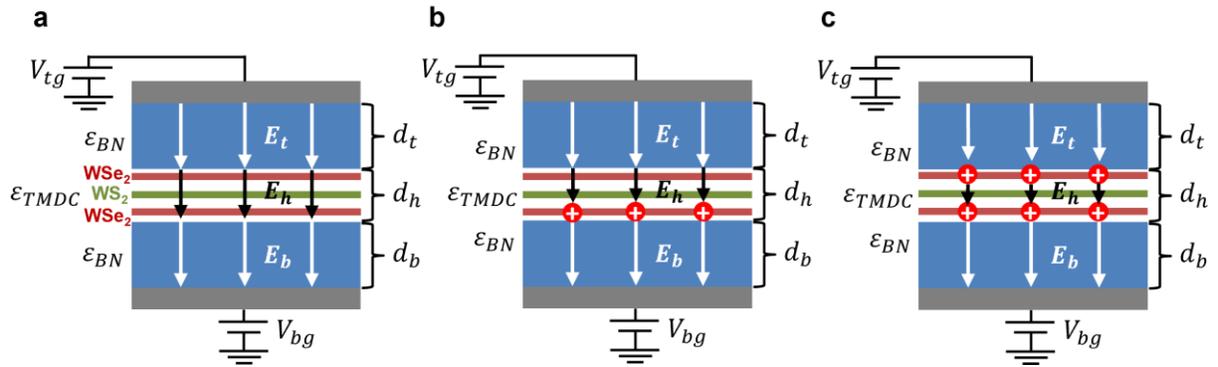

**Figure S14.** Schematics of the charge distribution in the trilayer region at n=0 (a), n=-1(b), and n=-2 (c).

**Case 1:** The heterostructure is in charge neutral region (n=0):

When the sample is insulating, there's a large potential drop on the quantum capacitance between the device and the ground[3,4]. Therefore, the heterostructure can be seen as floated. Considering the boundary conditions at the boundaries of dielectrics, we obtain the following equations:

$$\varepsilon_{BN}E_t = \varepsilon_{TMDC}E_h \qquad (11.1)$$

$$\varepsilon_{BN}E_b = \varepsilon_{TMDC}E_h \qquad (11.2)$$

Considering the potential drop between the top gate and the bottom gate, we have:

$$E_t d_t + E_h d_h + E_b d_b \approx E_t d_t + E_b d_b = V_{tg} - V_{bg} \qquad (11.3)$$

Here we neglected the potential drop in the TMDC heterostructure because its thickness is significantly smaller than the thickness of the BN layers.

From the above equations, we obtained the electric field inside the heterostructure:

$$E_h = \frac{\varepsilon_{BN}}{\varepsilon_{TMDC}} \frac{V_{tg} - V_{bg}}{2d} \tag{11.4}$$

**Case 2:** n = -1

Then we consider the threshold electric field to make one of the bilayers at n = -1 and the other at n = 0. In addition to the applied field between top gate and bottom gate, we need to consider the electric field generated by correlated holes. Let us consider the case where $V_{tg} > V_{bg}$ and holes only occupy the bottom WSe$_2$ layer. The boundary conditions become:

$$\varepsilon_{BN} E_t = \varepsilon_{TMDC} E_h \tag{11.5}$$

$$\varepsilon_{BN} E_b - \varepsilon_{TMDC} E_h = n_0 e \tag{11.6}$$

where $n_0 = 2.1 \times 10^{12} cm^{-2}$ is hole density corresponding to one hole per unit cell and e is the elementary charge. Therefore, we obtain:

$$E_h = \frac{\varepsilon_{BN}}{\varepsilon_{TMDC}} \frac{V_{tg} - V_{bg}}{2d} - \frac{n_0 e}{2\varepsilon_{TMDC}} = \frac{\varepsilon_{BN}}{\varepsilon_{TMDC}} E_{ext} - \frac{n_0 e}{2\varepsilon_{TMDC}} \tag{11.7}$$

where $E_{ext}$ is the external electric field defined in the Method Section of the main text. In our calculation we use dielectric constant values $\varepsilon_{BN} = 3.9\varepsilon_0$ and $\varepsilon_{TMDC} = 7.2\varepsilon_0$ [4]. Considering the critical external field of 44 mV/nm obtained from Fig. 4b, we obtain $E_h = -2.5\ mV/nm \approx 0.00V/nm$

**Case 3:** n=-2

At n=-2, there are symmetric amounts of holes occupying both the top WSe$_2$ layer and the bottom WSe$_2$ layer. The boundary conditions are:

$$\varepsilon_{TMDC} E_h - \varepsilon_{BN} E_t = n_0 e \tag{11.8}$$

$$\varepsilon_{BN} E_b - \varepsilon_{TMDC} E_h = n_0 e \tag{11.9}$$

Hence, we obtain:

$$E_h = \frac{\varepsilon_{BN}}{\varepsilon_{TMDC}} \frac{V_{tg} - V_{bg}}{2d} = \frac{\varepsilon_{BN}}{\varepsilon_{TMDC}} E_{ext} \tag{11.10}$$

From Fig. 4c, the critical external electric field to make the peak shift at n=-2 disappear is ~ 32.2 mV/nm. Therefore, we have $E_h = 17.4\ mV/nm$ in this case. According to the interlayer distance of about 1.2 nm we extracted from the dipolar excitons fitting (Supplementary Section 9), we obtain the energy difference between the two WSe$_2$ layers induced by the electrostatic potential at the critical external electric field to be about 20 meV.

## Supplementary section 12: Electric field dependence and doping dependence of PL spectra measured from device D3

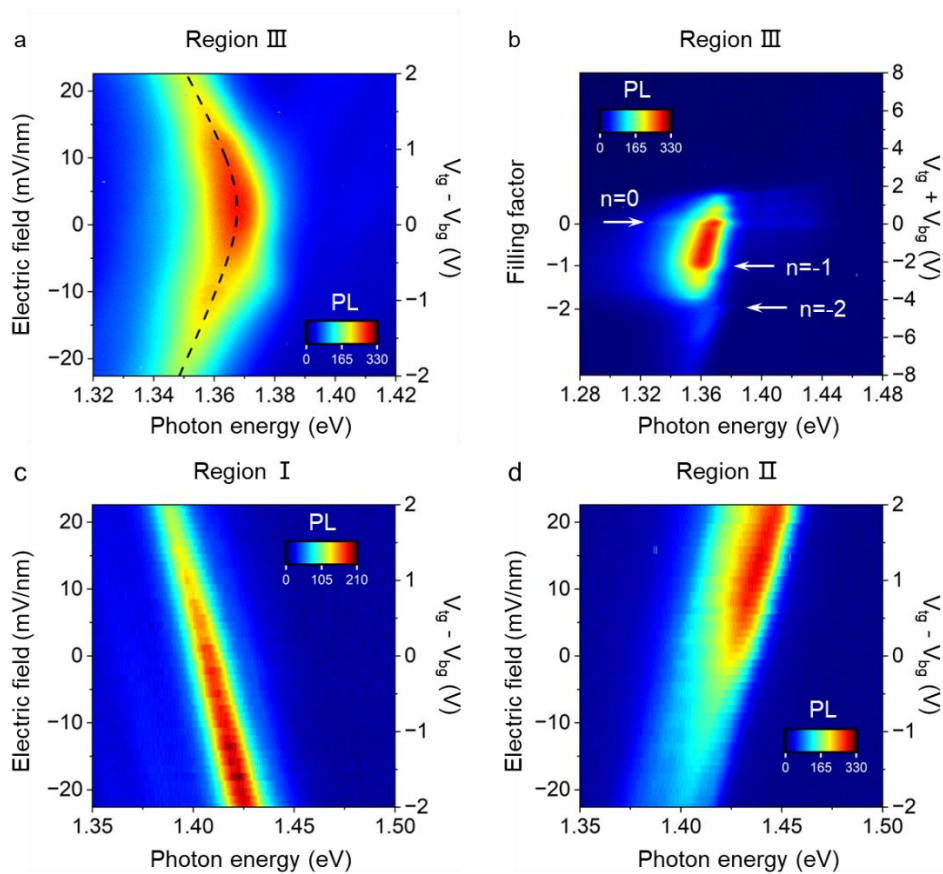

**Figure S15.** Electric field dependent and doping dependent PL spectra of another device D3. The data are taken with a 10 μW CW excitation with phonon energy centered at 1.959 eV and a temperature of 6 K.

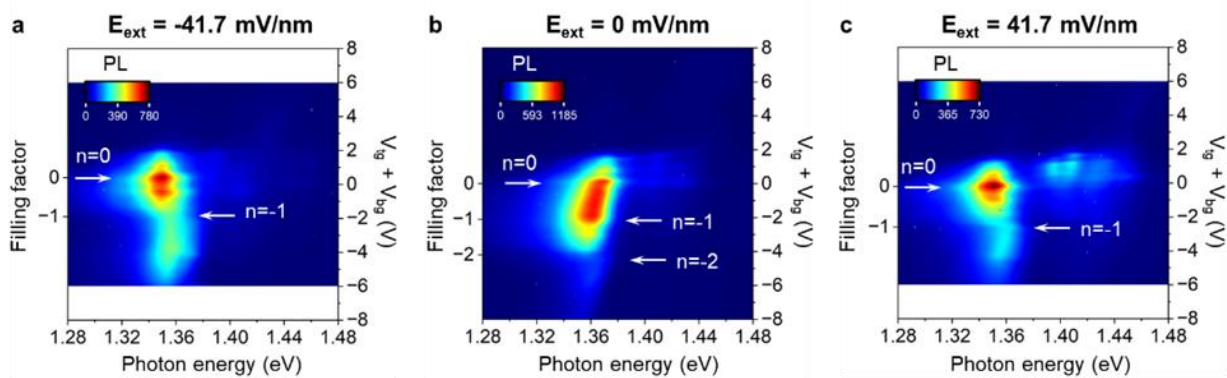

**Figure S16. Doping-dependent PL spectra from device D3 at different electric fields.** (a), (b) and (c) show the PL spectra taken from region III of device D3 with an out-of-plane external electric field of -41.7 mV/nm, 0 mV/nm and 41.7 mV/nm, respectively. The PL spectra are taken at 6K, with 10 µW 1.959 CW excitation.

## Supplementary section 13: Power dependence of PL spectra of device D3

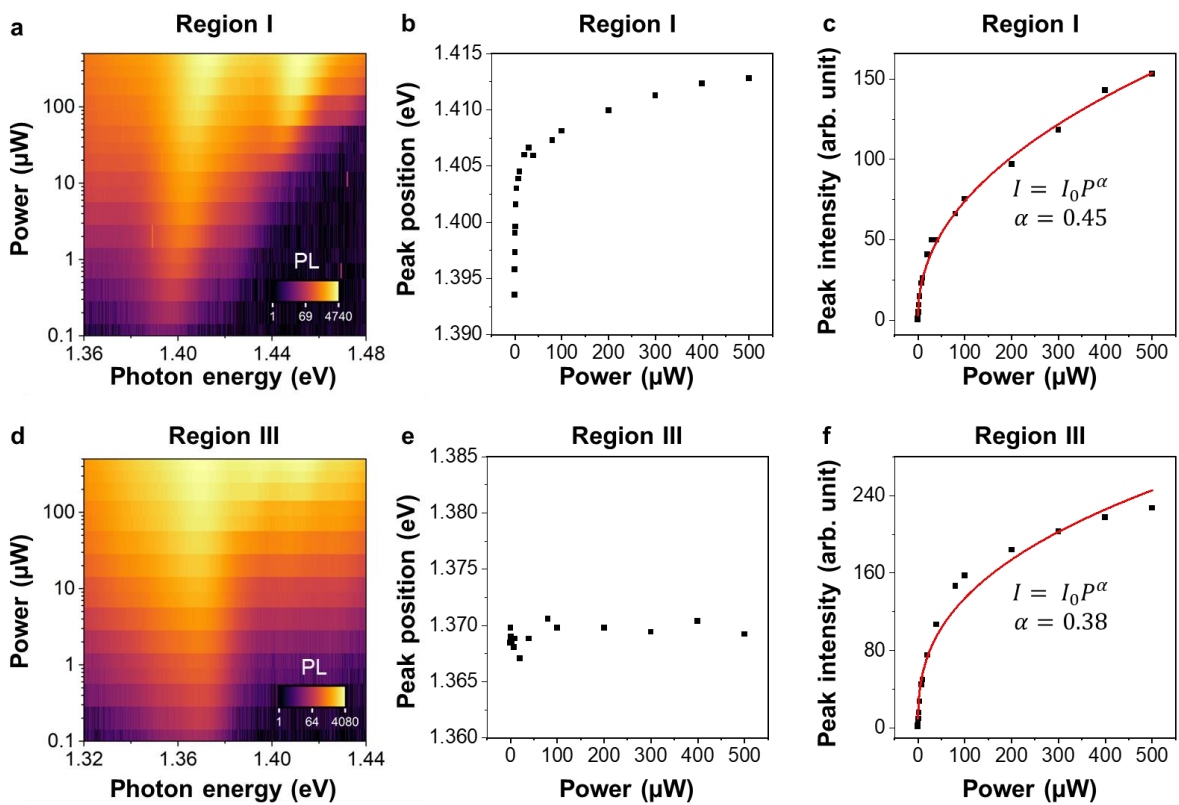

**Figure S17.** Excitation power dependence of PL spectra for region I and III of device D3.

## Supplementary section 14: Fitting to the two-level hybridization model for device D1 and D3

Here we fit the electric field dependence of quadrupolar exciton PL peak positions from device D1 and D3 to the model described in supplementary section 9. For device D1, we obtain $\Delta = 9 \pm 2$ meV, $X_1(0) = 1.368 \pm 0.001$ eV and $X_2(0) = 1.365 \pm 0.001$ eV. For device D3, the fitting yields $\Delta = 9 \pm 1$ meV, $X_1(0) = 1.380 \pm 0.001$ eV and $X_2(0) = 1.373 \pm 0.001$ eV.

Although the two dipolar excitons are of slightly different energies in device D3, they still form quadrupolar excitons due to the electric field tunability (schematically shown in Fig. S19). We note that the schematic shown in Fig. S19 also offers a way to experimentally determine whether the two dipolar excitons are of similar energies in the trilayer region: whether the quadrupolar excitons energy maximum, indicated by the arrow in Fig. S19, is at zero electric field or finite electric field. It is thus clear that the two dipolar excitons are of similar energies in device D1 (Fig. S18a) but slightly different in device D3 (Fig. S18b), consistent with the fitting results.

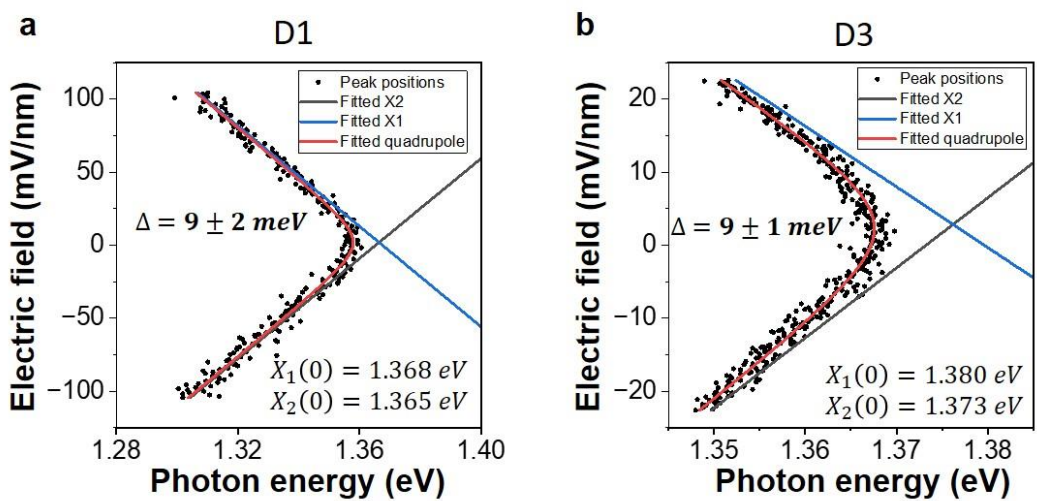

**Figure S18.** Fitted dipolar exciton and quadrupolar exciton energies obtained by setting the dipole moments and the peak positions at zero electric field of the two dipolar excitons as independent parameters from device D1 and D3.

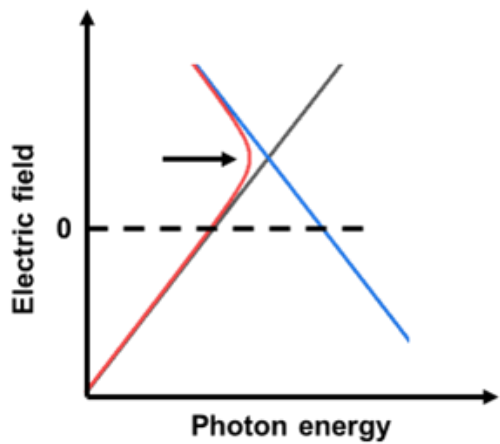

**Figure S19.** A schematic showing the hybridization of dipolar excitons with different energies at zero electric field.

We have observed the signs of interlayer quadrupolar exciton PL in 19 devices, and 9 of them can be well fitted with the quadratic electric field dependence. Among them, the interlayer Mott insulator behavior presented in the main text is observed in 6 devices. We suspect that the observation of the correlated states is more demanding on the relative angle alignment between the top and bottom $WSe_2$ layer, which is needed to ensure the same moiré supercell for the top and bottom moiré bilayer.

**Supplementary section 15: PL spectra of device D1 measured with 1.698 eV laser excitation**

Here we show the PL spectra of device D1 with the exciton photon energy of 1.698 eV, which is in resonance with the moiré A exciton of $WSe_2$. The spectra are similar to what we show in the main text and supplementary section 4. Both excitation photon energy of 1.959 eV and 1.698 eV are much larger than the interlayer exciton resonance of ~1.4 eV.

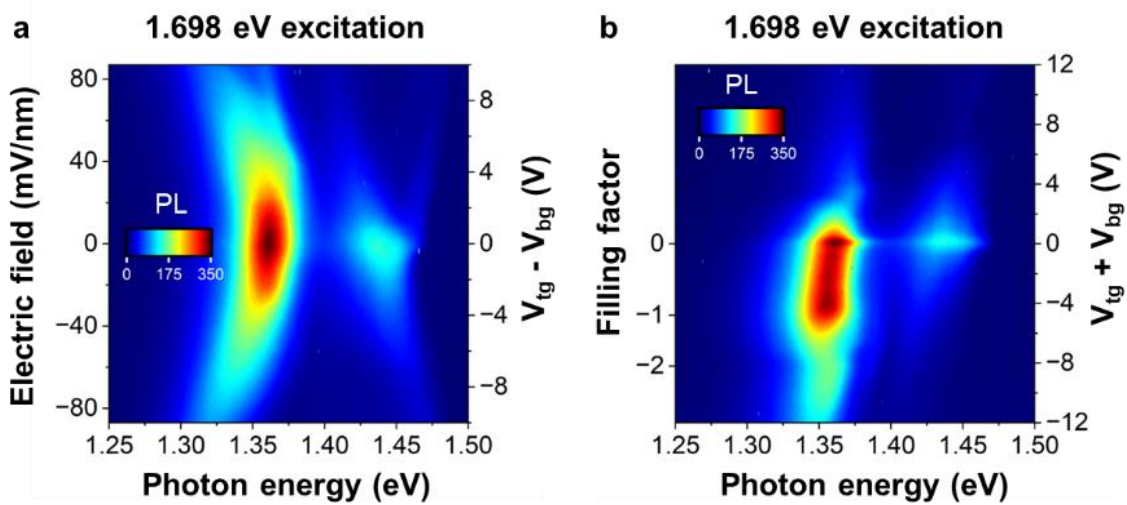

**Figure S20. Electric field dependence and doping dependence of the PL spectra from region III of device D1.** CW laser with the phonon energy centered at 1.698 eV, in resonance with the moiré A exciton of $WSe_2$, is used as the excitation. The excitation power is 200 µW, and the data are taken at a temperature of 6 K.

### Supplementary section 16: Estimation of exciton density

Here we derive the exciton density as a function of excitation power following the method described in ref.5. We consider the reflection and absorption of each layer of material. The effective power density at the heterobilayer is expressed as:

$$P_{eff} = P(1 - R_G - A_G)(1 - R_{BN})(1 - R_{TMDC}) \qquad (16.1)$$

Where $A_G$ and $A_{TMDC}$ denote the absorptance of the top graphite and the top BN, respectively. $R_G$, $R_{BN}$ and $R_{TMDC}$ denote the reflectance of the top graphite, top BN and the heterobilayer. $P$ is the power density at the surface of the sample, which can be calculated using the excitation power and the beam spot size d = 2 µm. For few-layer graphene top gate, the absorptance $A_G$ is negligible. At each dielectric interface, the reflectance R can be calculated by:

$$R(\lambda) = \left| \frac{n_R(\lambda) - n_T(\lambda)}{n_R(\lambda) + n_T(\lambda)} \right|^2 \qquad (16.2)$$

Where $n_R(\lambda)$ and $n_T(\lambda)$ are the real refractive indexes at the reflectance side and the transmittance side, respectively.

Using the refractive index values adapted from previous works[6-9], we estimate $P_{eff} = 0.64P$ for 1.96 eV excitation.

We estimate the proportion of the power absorbed by the hetero-trilayer using the values in ref.6, which is 0% for $WS_2$ and 2% for $WSe_2$ at 1.96 eV. For region III, assuming 4% absorption and unitary conversion efficiency, we obtain the exciton generation rate g = $2.6 \times 10^{18}\ cm^{-2} \cdot s^{-1}$ for 1 µW 1.96 eV excitation. Using a measured exciton lifetime around 19 ns, the steady-state exciton density can be estimated by $d_{ex} = g \cdot \tau$ which is $5.0 \times 10^{10} cm^{-2}$ for 1 µW excitation. With 200 µW 1.96 eV excitation, the steady-state exciton density is $1.0 \times 10^{13} cm^{-2}$ for 200 µW 1.96 eV excitation and $2.5 \times 10^{12} cm^{-2}$ for 50 µW 1.96 eV. Similarly, for region II, 1 µW excitation power corresponds to an exciton density of $2.7 \times 10^{10} cm^{-2}$.

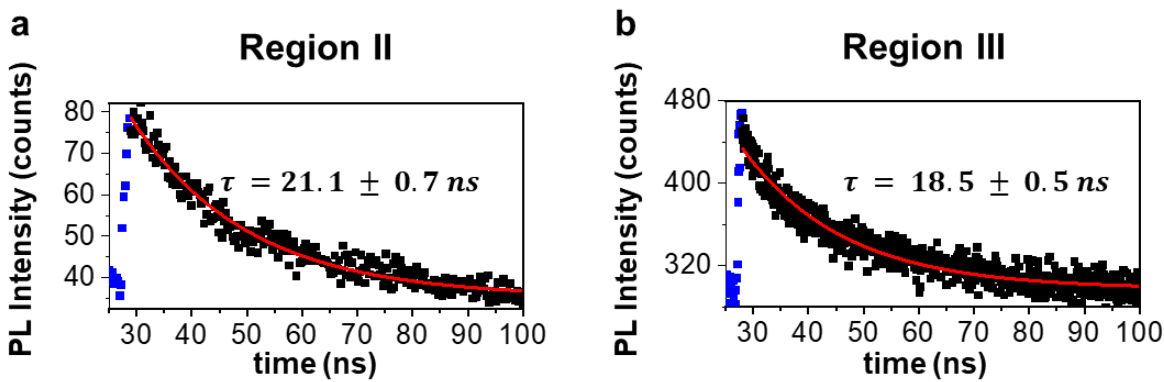

**Figure S21. Time-resolved photoluminescence (TRPL) measured from region II and region III.** (a) and (b) are the TRPL data taken from region II and region III of device D1, respectively. The red lines indicate fitting results using an exponential decay function $I(t) = I_0 e^{(t-t_0)/\tau}$. The TRPL measurements were performed using a pulsed laser centered at 1.771 eV with a repetition rate of 10 MHz. The excitation power is 400 μW.

## Supplementary section 17: Discussion of the energy scales related to the interlayer Mott state

The n=-2 state in Fig. 3 is the case where the top and bottom moiré superlattices are half-filled by holes (-1 for the top $WSe_2/WS_2$ and -1 for the bottom $WS_2/WSe_2$ moiré superlattice). Therefore, the bandgap will be the smaller onsite repulsion energy (U) of the top and bottom Mott insulator minus the energy $\Delta^{\pm}$. In Fig. S22, we assume the same U for both valence bands for simplicity. We believe that this U could be more than 50 meV (manuscript under preparation and a recent work on arXiv[10]).

The n=-1 state in Fig. 3 is the newly discovered hybridized Mott insulator state. It can be understood schematically as shown in Fig. S22. The hybridization of the flat valance band gives to the bonding state (share the origin of the quadrupolar excitons) and antibonding state, separated by the energy gap $\Delta^{\pm}$. For the n=-1 state, the bonding state will be half filled as the LHB, while the UHB of the bonding state will be at the energy U higher. As we extracted $\Delta$ to be 9 meV (SI Section 14) for device D1, which corresponding to a $\Delta^{\pm}$ of 18 meV, the UHB will be higher than the antibonding state, so that the next added electron will occupy the antibonding state, leading to the bandgap of $\Delta^{\pm}$. As $\Delta^{\pm}$ is smaller than U-$\Delta^{\pm}$, we would expect the shift at n=-1 to be smaller than that at the n=-2.

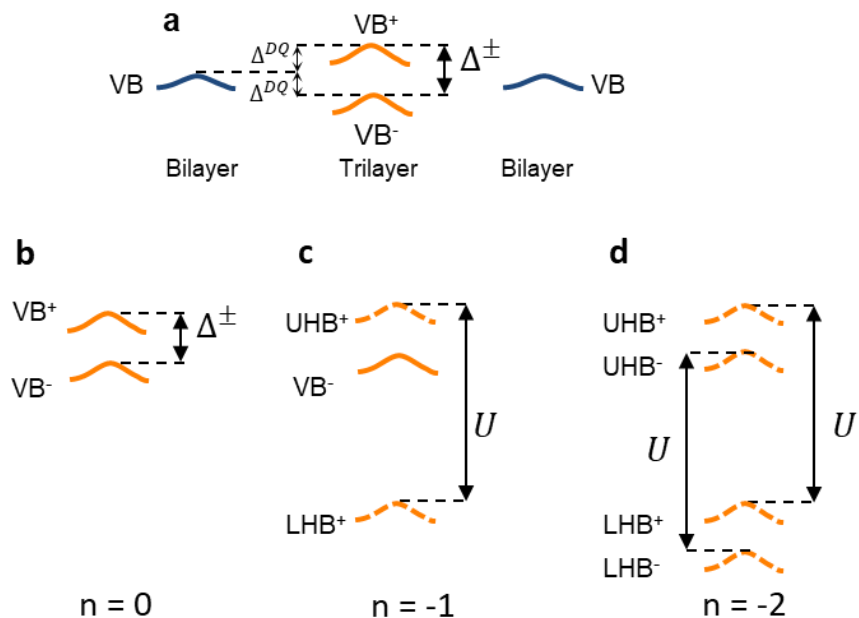

**Figure S22. Schematics showing the interlayer hybridized Mott insulating state.** (a) shows the valence band hybridization process in WSe$_2$/WS$_2$/WSe$_2$ trilayers. (b), (c) and (d) show the schematics of the bandstructure in the trilayer region at hole filling of n=0, n=-1 and n=-2, respectively

**Supplementary section 18: Discussion of the nature of the high-energy peak from device D1**

The high-energy PL peaks in Fig. 3 and Fig. S4 occur at high excitation power and are not universal among all the devices we have studied. It could be attributed to other exciton modes at higher energy, including possible quadrupolar exciton modes from hybridization of higher energy dipolar excitons. We do not believe that it is from the high energy branch (antisymmetric mode associated with the symmetric quadrupolar exciton mode discussed in the main text) quadrupolar exciton (because of the large energy separation from the ground state quadrupolar exciton. It is likely that the simplified model did not consider moiré effects or possible complications from other dipolar excitons (such as spin-singlet exciton[11]). As the PL spectra are most sensitive to the ground state, we focus on the symmetric quadrupolar exciton branch (lower energy branch) in this work and leave the investigation of the high energy mode(s) to future exploration.

**Supplementary section 19: PL spectra from a WS$_2$/WSe$_2$/WS$_2$ heterostructure**

Fig.S23 shows the doping-dependent and electric-field-dependent of PL spectra measured from a WS$_2$/WSe$_2$/WS$_2$ device (D4).

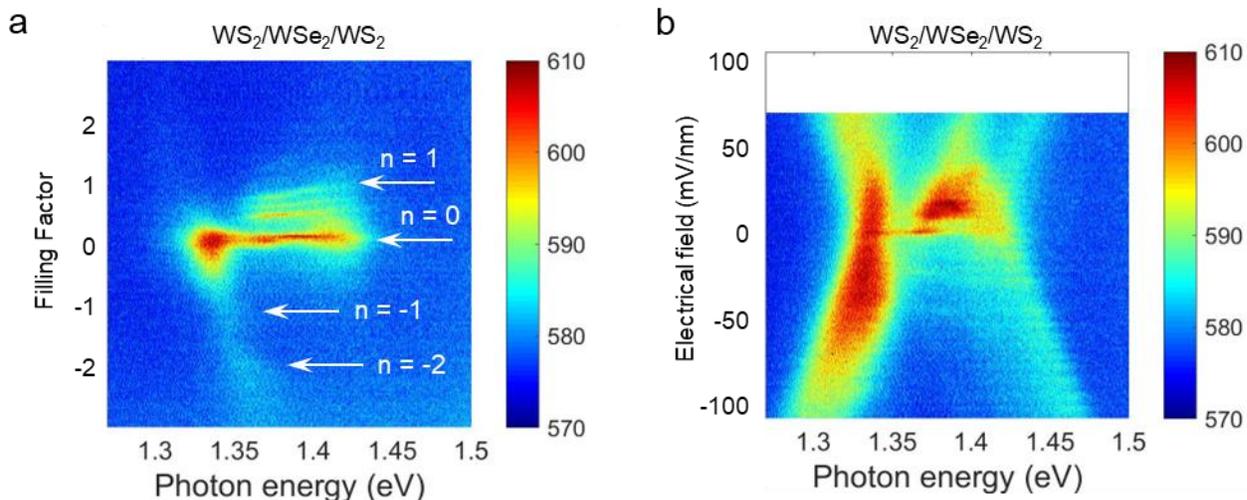

**Fig.S23. PL spectra from a WS$_2$/WSe$_2$/WS$_2$ heterostructure.** (a) and (b) show the doping dependence and the electric field dependence of PL spectra from WS$_2$/WSe$_2$/WS$_2$ device D4.

## References

1.  Regan, E. C. *et al.* Mott and generalized Wigner crystal states in $WSe_2/WS_2$ moiré superlattices. *Nature* **579**, 359–363 (2020).

2.  Tang, Y. *et al.* Simulation of Hubbard model physics in $WSe_2/WS_2$ moiré superlattices. *Nature* **579**, 353–358 (2020).

3.  Zhang, Y. *et al.* Direct observation of a widely tunable bandgap in bilayer graphene. *Nature* **459**, 820–823 (2009).

4.  Jauregui, L. A. *et al.* Electrical control of interlayer exciton dynamics in atomically thin heterostructures. *Science* **366**, 870–875 (2019).

5   Bai, Y., *et al.* Evidence for exciton crystals in a 2D semiconductor heterotrilayer. arXiv:2207.09601 (2022).

6   Li, Y. *et al.* Measurement of the optical dielectric function of monolayer transition-metal dichalcogenides: $MoS_2$, $MoSe_2$, $WS_2$, and $WSe_2$. *Physical Review B* **90**, 205422 (2014).

7.  Lee, S.-Y. *et al.* Refractive index dispersion of hexagonal boron nitride in the visible and near-infrared. *Phys. Status Solidi B* **256**, 1800417 (2019).

8.  Wang, X. *et al.* Strong anomalous optical dispersion of graphene: complex refractive index measured by picometrology. *Optics Express* **16**, 22105-22112 (2008).

9.  Hsu, C. *et al.* Thickness-dependent refractive index of 1L, 2L, and 3L $MoS_2$, $MoSe_2$, $WS_2$, and $WSe_2$. *Advanced optical materials* **7**, 1900239 (2019).

10. Li, H. *et al.* Mapping Charge Excitations in Generalized Wigner Crystals. arXiv:2209.12830v2

11. Wang, T. *et al.* Giant Valley-Zeeman Splitting from Spin-Singlet and Spin-Triplet Interlayer Excitons in $WSe_2/MoSe_2$ Heterostructure. *Nano Lett.* **20**, 694–700 (2020).